\documentclass[12pt, draftclsnofoot, onecolumn]{IEEEtran}
\usepackage{cite}
\usepackage{amsmath,amssymb,amsfonts}
\usepackage{algorithmic}
\usepackage{graphicx}
\usepackage{textcomp}
\usepackage{xcolor}
\usepackage{algorithm}
\usepackage{graphicx}
\usepackage{subfig}
 \pdfoutput=1

\begin{document}
	
	\title{\huge \textcolor{blue}{Resource Management for Transmit Power Minimization in UAV-Assisted RIS HetNets Supported by
		Dual Connectivity}}
	\author{  Ata Khalili, \textit{Member, IEEE}, Ehsan Mohammadi Monfard, Shayan Zargari, Mohammad~Reza.~Javan,  \textit{Senior Member, IEEE}, Nader~Mokari,  \textit{Senior  Member, IEEE}, and Eduard A. Jorswieck,  \textit{Fellow, IEEE}
		\thanks{A. Khalili,~E.~Monfared, Sh. Zargari, and N. Mokari are with the Department of Electrical and Computer Engineering,~Tarbiat Modares University,~Tehran,~Iran, (Dr. Mokari e-mail: nader.mokari@modares.ac.ir). M.~R.~Javan is with the Department of Electrical Engineering, Shahrood University of Technology, Iran, (e-mail: javan@shahroodut.ac.ir).~Eduard A. Jorswieck is with the Institute for Communications Technology, TU
			Braunschweig, Germany, E-mail: jorswieck@ifn.ing.tu-bs.de.~This work was supported by the joint Iran national science foundation (INSF) and German research foundation (DFG) under grant No. 96007867.}}
	\maketitle
	\begin{abstract}
	This paper proposes a novel approach to improve the performance of a heterogeneous network (HetNet) supported by dual connectivity (DC) by adopting multiple unmanned aerial vehicles (UAVs) as passive relays that carry reconfigurable intelligent surfaces (RISs) .~\textcolor{blue}{More specifically, RISs are deployed under the UAVs termed as UAVs-RISs that operate over the micro-wave (µW) channel in the sky to sustain a strong line-of-sight (LoS) connection with the ground users.~The macro-cell operates over the µW channel based on orthogonal multiple access (OMA),~while small base stations (SBSs) operate over the millimeter-wave (mmW) channel based on non-orthogonal multiple access (NOMA)}. We study the problem of total transmit power minimization by jointly optimizing the trajectory/velocity of each UAV, RISs' phase shifts, subcarrier allocations, and active beamformers at each BS.~The underlying problem is highly non-convex and the global optimal solution is intractable.~To handle it, we decompose the original problem into two subproblems,~i.e.,~a subproblem which deals with the UAVs' trajectories/velocities, RISs' phase shifts, and subcarrier allocations for µW; and a subproblem for active beamforming design and subcarrier allocation for mmW.~In particular,~we solve the first subproblem via the \textcolor{blue}{dueling deep Q-Network (DQN) learning approach by developing a distributed algorithm which leads to a better policy evaluation.} Then,~we solve the active beamforming design and subcarrier allocation for the mmW via the successive convex approximation (SCA) method.~\textcolor{blue}{Simulation results exhibit the effectiveness of the proposed resource allocation scheme compared to other baseline schemes. In particular, it is revealed that by deploying UAVs-RISs, the transmit power can be reduced by $6$ dBm while maintaining similar guaranteed QoS.}
	\end{abstract}

	\begin{IEEEkeywords}
		Unmanned aerial vehicle (UAV), reconfigurable intelligent surface (RIS), non-orthogonal multiple access (NOMA), deep Q-network (DQN) learning.
	\end{IEEEkeywords}

	\IEEEpeerreviewmaketitle

	\section{Introduction}
	\subsection{Motivation}
	The next generation of wireless communication systems operate at the millimeter-wave (mmW) frequency to satisfy the growing demand for high capacity and spectral efficiency (SE) \cite{6G}. Although the mmW frequency introduces many difficulties, such as sensitivity to the blockage, adopting such bands can significantly improve the throughput of the wireless communication systems by promoting the utilization of spectrum, especially for the small base stations (SBSs) of heterogeneous networks (HetNets) \cite{Peng}. In particular, the HetNets can offload data from the macro base stations (MBSs) to SBSs, which extends the wireless coverage area and decreases the transmit power.~In other words, each user can simultaneously receive data from two different frequencies, i.e., micro-wave (µW) and mmW links
	%This connectivity comes from adopting various frequencies
	\cite{Moltafet}.~Besides, the SBSs can reuse and share the same channels with the MBSs to achieve higher SE. Accordingly, the HetNets are considered as an excellent strategy for improving the throughput of the cellular networks\cite{Zhaoo}. However, data traffic in such networks is growing rapidly. To address this issue, the power domain non-orthogonal multiple access (PD-NOMA) is introduced as a nominee technique in rendering efficient utilization of the spectrum \cite{Liu,Saito}.~In PD-NOMA schemes, users can be multiplexed on the same shared frequency-time resource blocks with different power levels \cite{Z. Yang}. In contrast to the conventional orthogonal multiple access (OMA) schemes, the users are ordered based on which users with higher order can apply the successive interference cancellation (SIC) technique to their decode data by excluding the interference signals from other users with the lower order. Thus, the PD-NOMA scheme offers better user fairness \cite{Ding}, as well as higher SE as compared to the traditional OMA schemes in HetNets \cite{Zhu}.
	
	On the other hand, unmanned aerial vehicles (UAVs), generally identified as aerial base stations (BSs), have drawn significant attention for different applications such as relaying, information gathering, and data distribution \cite{Shii}.
	%since they are provided with high-level transceivers and batteries \cite{Shii}.
	%Setting up safe line-of-sight (LoS) communication links with ground users has always been critical issue\cite{Limm}.~Accordingly,~
	Deploying UAVs as flying BSs provides a better chance to create line-of-sight (LoS) links through its high plasticity and operability, which can enhance the performance of the current terrestrial wireless communication systems\cite{Limm}.~However, one eminent difficulty of the UAV is blockages (such as trees and buildings) that can either block transmit signals or diminish their power \cite{Kandeepan}. Also, long-range wireless communication between UAV-users results in large path-loss that cannot be neglected \cite{Alouinii}.
	
%~Hence,~it might lead to poor connection and coverage limitation\cite{Kandeepan}. Accordingly, long-range wireless communication between each UAV and ground users results in large path-loss that cannot be neglected in UAV networks \cite{Alouinii}.
	
	At the same time, recent developments on “meta-surface” allow the implementation of the reconfigurable intelligent surfaces (RISs) \cite{Bennis1}, which are excellent candidates to overcome the difficulties mentioned above. In particular, conventional relays based wireless systems operate in either decode-and-forward (DF) or amplify-and-forward (AF) modes,~i.e.,~they can not receive and transmit information simultaneously. However, RIS operates in the full-duplex (FD) mode without suffering from any self-interference (SI) and additional receiving noise, which is very pleasant from the perspective of the practical implementation\cite{Wu1}. More specifically, RIS comprises many passive reflecting components with reconfigurable phase shifts and small energy consumption,
	%since it needs neither decode nor amplifies the incident signals.
	which can produce a favorable wireless propagation environment \cite{shayan,kawn}.~Hence,~it can be regarded as a potential and novel solution to overcome hardware expenses, complexity, and energy consumption in UAV communication systems. 
	%In particular, multiple RISs can be deployed on various UAVs termed as UAVs-RISs in the sky to sustain a strong connection with users and improve reliability by establishing several LoS connections.
	
%	Due to the rapid growth of data traffic, the power domain non-orthogonal multiple access (PD-NOMA) is introduced as a nominee technique in rendering efficient utilization of spectrum\cite{Liu,Saito}.~The PD-NOMA scheme exploits the power domain in which users can be multiplexed on the same shared frequency-time resource blocks with different power levels\cite{Z. Yang}. In contrast to the conventional orthogonal multiple access (OMA) schemes, the user with the better SINR can apply the successive interference cancellation (SIC) technique to decode its data by excluding the interference signals from other users with the worse SINR. Thus, the PD-NOMA scheme offers better user fairness \cite{Ding}, as well as higher SE as compared to the traditional OMA schemes in HetNets\cite{Zhu}.

	\subsection{Related Works}
	The related works on this topic can be classified into four groups, namely: 1) PD-NOMA-based systems with HetNet; 2) RIS-based systems; 3) UAV-based systems 4) UAV-RIS-based systems.
	
	\textit{1) PD-NOMA-based systems with HetNet:} The PD-NOMA is a suitable candidate for multiple access, which can be exploited in HetNets to improve the SE. For instance, the authors in \cite{Moltafet} investigate the impact of the PD-NOMA scheme in a HetNet, where energy efficiency (EE) is maximized with the existence of both the mmW and µW links in both access and fronthaul links. The authors in \cite{Qian} consider dual connectivity (DC) to simplify flexible traffic offloading
	%, where each user can interact with the MBS and offload traffic through an SC under the PD-NOMA scheme. 
	In particular, the total power dissipation minimization of the SBSs and MBSs is studied under the PD-NOMA scheme. However, \cite{Moltafet} and \cite{Qian} have mainly focused on a single-carrier NOMA system. In the multi-carrier (MC)-NOMA systems, each subcarrier can be assigned to two or more users with various channel conditions. The deployment of the PD-NOMA in a multi-carrier (MC) system is studied in \cite{Yuan,Wei,Sun}, which reveal that fairness and the spectrum usage can be improved significantly. A suboptimal power and subcarrier allocation algorithm for a MC-NOMA system is proposed in \cite{Yuan}, where the weighted sum rate (WSR) is maximized. The optimal algorithms to minimize the transmit power of the BS and maximize the WSR are studied in \cite{Wei} and \cite{Sun}, respectively.

	\textit{2) RIS-based systems:} Several recent works have studied the application and performance of the RIS in wireless communication systems. For instance, in \cite{Alouini}, a minimum mean squared error-based channel estimation is proposed to design a RIS-aided multi-user multiple-input single-output (MISO) system. The authors in \cite{Zappone} study the EE, where the transmit power and phase shifts at the BS and RIS, respectively, are optimized. In \cite{Zhang}, the total transmit power at the multi-antenna access point (AP) is minimized via jointly designing the transmit beamformers at the AP and passive components at the RIS. %Besides, a large number of reflecting components are taken into consideration to investigate the asymptotic performance.
	The authors in \cite{Nadeem} study the RIS-aided MISO downlink (DL) wireless communication system, where the minimum signal-to-interference-plus-noise ratio (SINR) is maximized. In \cite{Hanzo}, a multi-cell scenario is considered where the RIS is located at the cell boundary to mitigate the inter-cell interference and help the DL transmission. In particular, the WSR is maximized by jointly designing the active precoding and the passive beamformers at the BS and RIS, respectively. A novel framework is proposed in \cite{Bennis1} to design a RIS-assisted mmW BS, where the optimal transmission precoder, power allocation, and reflective components are obtained to maximize the average sum-rate under perfect and imperfect channel state information (CSI). However, in general, practical RISs have a large number of components, which renders continuous adjustment of the phase shifts costly and even impossible\cite{Wu1}. Hence, it is more cost-effective to deploy discrete phase shifts at the RIS. For this reason, the authors in \cite{Zhang1} propose discrete phase shifts to minimize the total transmit power.

	The application of the PD-NOMA in the RIS systems is investigated in \cite{Shi2,Zheng24,Ding56,Hanzo1}. The power minimization problem is studied in \cite{Shi2}, where the beamforming vectors at the BS and the phase shifts at the RIS are jointly optimized. A comparison between PD-NOMA and OMA from the theoretical aspect is studied in \cite{Zheng24}. In particular, the transmit power of the BS is minimized which unveils that for users near to the RIS, the PD-NOMA scheme may perform worse than time division multiple access (TDMA). The authors in \cite{Ding56} propose a simple RIS-NOMA system where spatial division multiple access (SDMA) is adopted at the BS to create orthogonal beams and then exploit PD-NOMA to serve extra cell-edge users on such beams. In \cite{Hanzo1},~a NOMA-based RIS system is considered where the outage probability, ergodic rates, SE and EE are all obtained in closed-form expressions.~\textcolor{blue}{The authors in \cite{R2} consider a wireless communication environment with distributed RISs where system EE is maximized via adjusting the on-off status of each RIS and optimizing the reflected phase shifts at the RISs.}
	
	\textit{3) UAV-based systems:}
	UAV is considered as a novel technology with the purpose of providing services for industrial applications such as transportation and relaying \cite{Gupta,Lim}. The deployment of the UAVs as flying BSs in favor of wireless communication range and capacity improvement is studied extensively in \cite{Debbah}.~In particular, cooperative services and coverage extension can be offered by UAVs due to the capability in mobility and hovering \cite{Feng}.~Especially, for cases where the channel has a bad condition due to the blockage or cell-edge position \cite{Lim1}. The minimization of the outage probability is considered in\cite{Bian} by jointly designing the UAV's trajectory as well as transmit power of the users and UAV. A DL UAV-based relay system under the PD-NOMA scheme is investigated in \cite{Wang}, where the UAV flies in a circular path and acts as a DF relay to minimize the maximum outage probability.
	
	\textit{4) UAV-RIS-based systems:}
	The UAV and RIS together have drawn significant research consideration for the deployment of the sixth-generation (6G) of the wireless communication networks.~The authors in \cite{Shafique} analyze the ergodic capacity, SNR outage probability, and EE under three distinct modes, i.e., only UAV, only RIS, and integrated UAV-RIS modes. In \cite{HuaHua}, the data rates of all RISs to a given BS are maximized by exploiting the UAV. More specifically, a maximization-minimization optimization problem is investigated by jointly optimizing the phase shifts at the RIS, UAV's trajectory, and RIS scheduling. The authors in \cite{WangWang} maximize the data rate and geographical fairness of all the users by jointly optimizing the trajectory of UAV and passive phase shifts at the RIS. In particular, two approaches based on the deep Q-network (DQN) and deep deterministic policy gradient (DDPG) are proposed for discrete and continuous scenarios, respectively. The authors in \cite{Hassan} study the disjoint UAV and RIS system, where the UAV assists in transmitting signals to the RIS as a result of the down-tilt of the BS’s antennas.
	
	On the other hand, the UAV-mounted RIS can further enhance the reliability of the communication systems and overcome the blockages by modifying its position and the phase shifts. In \cite{Cai}, UAV-carried RIS is considered to minimize the average total power dissipation of the system by jointly optimizing the UAV's trajectory/velocity and phase shifts at the RIS. In \cite{Renzo}, the average achievable rate maximization problem for an UAV-carried RIS is studied by jointly optimizing UAV's trajectory and passive elements at the RIS. The authors in \cite{Saad} consider an UAV-RIS as an energy harvester, where the capacity of the system is maximized by adopting an reinforcement learning (RL) approach. In \cite{Jiao}, an UAV-carried RIS aided NOMA scheme is investigated where the rate of the strong user is maximized while guaranteeing the rate of the weak user. In particular, the location of the UAV-RIS is optimized first and then the transmit beamformers and phase shift of the BS and the UAV-RIS are optimized alternatively.~\textcolor{blue}{In \cite{R1}, the secure EE maximization problem for a UAV-RIS system is investigated where phase shift at the IRS, UAV’s trajectory, user association, and transmit power are jointly optimized. In particular, a three-phase iterative algorithm based on the SCA method is proposed to address the optimization problem.}

	%In particular, we focus on the DRL approach to address resource allocation in a HetNet assisted UAV-RIS.
	%In particular, we would like to see whether employing UAV-aided RIS can be beneficial in an MBS and how the performance of such a network is compared to a conventional one that does not exploit any UAV-RIS.
	%    \cite{Hassan}-\cite{Saad}.

	\subsection{Contributions}

	\textcolor{blue}{The resource allocation design for MC-PD-NOMA HetNets enabling DC with the help of the UAVs-RISs has not been studied in the literature to the best of the authors' knowledge. Thus, it motivates us to study the potential benefits of multiple UAV-mounted RIS in a HetNet for establishing an energy-efficient network. In this paper, we investigate whether employing UAVs-RISs can be beneficial in an HetNet and how such a network operates compared to a conventional one that does not exploit any UAV-RIS.}
	
	\textcolor{blue}{$\bullet$ We investigate the potential deployment of multiple UAVs-RISs in a HetNet to help users with the non-line-of-sight (NLoS) links or poor channel conditions. Specifically, each UAV-RIS can modify its position and phase shifts to enhance the system performance in terms of transmit power and network coverage.}

	\textcolor{blue}{$\bullet$ To assess the performance, a total transmit power minimization problem is formulated by jointly optimizing UAVs' trajectory/velocity, subcarrier allocations for macro-cell and small-cells, RIS phase shifts, and active beamformers at the MBS and the SBSs. }
	
	\textcolor{blue}{$\bullet$ To solve this complicated problem, the main problem is divided into two sub-problems: trajectory/velocity, phase shifts, and subcarrier allocation for the $\mu$W optimization subproblem (named as higher level since the dimension of the problem is high) and joint beamforming design and subcarrier allocation for mmW sub-problem (named as lower level).}
	
	\textcolor{blue}{$\bullet$  For the higher level sub-problem, based on the deep Reinforcement Learning (DRL) technique, a dueling deep Q-Network (DQN) [50] learning problem is used to train and update the trajectories/velocities of the UAVs, phase shifts at the RISs, and subcarrier allocation for $\mu$W. For the lower level, the joint beamforming design and subcarrier allocation for mmW are handled by exploiting the semi-definite relaxation (SDR) technique, majorization-minimization (MM) algorithm, and penalty function method to form a reward function emerging from taking actions in the higher level.}
	
	\textcolor{blue}{$\bullet$ Simulation results exhibit the effectiveness of the proposed resource allocation scheme for the considered network with multiple UAVs-RISs under the MC-PD-NOMA scheme. They also reveal their superior performance in terms of the total transmit power compared to other baseline schemes.}

	$\mathbf{Notations:}$ Column vectors and matrices are represented by boldfaced lowercase and uppercase letters, e.g., $\mathbf{a}$ and $\mathbf{A}$, respectively. The superscript $(\cdot)^T$ and $(\cdot)^H$ denote the transpose and Hermitian conjugate transpose, respectively. $\text{Tr}(\mathbf{A})$ and $\text{Rank}(\mathbf{A})$ denote the trace and the rank of matrix $\mathbf{A}$, respectively. $\mathbf{A}\succeq\mathbf{0}$ indicates a positive semidefinite matrix. $\text{diag}(\cdot)$ is the diagonalization operation. The distribution of a circularly symmetric complex Gaussian (CSCG) random vector with mean $\boldsymbol{\mu}$ and covariance matrix $\mathbf{C}$ is denoted by $\sim \mathcal{C}\mathcal{N}(\boldsymbol{\mu},\,\mathbf{C})$. $\mathbb{E}(\cdot)$ is the statistical expectation, and $\mathbb{C}^{M\times N}$ represents an $M\times N$ dimensional complex matrices. $\mathbb{R}^{N\times 1}$ denotes an $N\times 1$ dimensional real vectors, and $\mathbb{N}$ is an real scalar. $\mathbf{I}_M$ denotes the $M$-dimensional identity
	matrix. $\|\mathbf{a}\|$ and $|b|$ denote the Euclidean norm of a vector and the magnitude of a complex number $b$, respectively. $\nabla_\mathbf{x}f(\mathbf{x})$ denotes the gradient vector of function $f(\mathbf{x})$ with respect to
	$\mathbf{x}$.

	\section{System Model and Problem Formulation}
	\subsection{System Model}
	In this paper, we consider a downlink (DL) HetNet serving macro-cell users (MUEs) and small-cell users (SUEs), where the SBSs operate on the mmW channel,~while the MBS operates on µW. More precisely, the SBSs and MBS occupy two different frequency bands where the MC-PD-NOMA and orthogonal frequency division multiple access (OFDMA) schemes are adopted at the SBSs and MBS, respectively. Since mmW operates on the ultra-high frequency band, the MC-PD-NOMA scheme can be exploited to enhance the performance of the mmW channel\footnote{Note that due to the high attenuation features of the mmW links, performing the PD-NOMA scheme improves the performance gain concerning SE and EE. Also, we consider OFDMA for the MBS since the implementation of the PD-NOMA scheme is quite challenging, and assume UAVs enabled narrow-band DL OFDMA wireless communication.}. Note that in this way, the interference between the SBSs and MBS can be eliminated due to the separated frequency bands of MBS and SBSs\footnote{Note that the spatial division multiple access (SDMA) scheme cannot perform properly for close cellular users in dense networks, particularly the ones who are located at the cell edge. This is mainly because the number of users in small-cells is large, and also due to the blockage of the mmW links, they lie in similar directions from their serving SBS, which imposes a high correlation among their channel links. These correlations diminish the achievable spatial multiplexing gain as a result of the more stringent co-channel interference. Thus, we consider the MC-PD-NOMA scheme to serve the users on different carriers with various power levels in order to overcome the co-channel interference.}. The set of BSs is denoted by $\mathcal{J}=\{0,1,2,..., J\}$, where index 0 indicates the MBS. All users are randomly distributed with uniform distribution in a coverage area of the MBS overlaid by the SBSs.~The sets of all users for MBS and SBSs are indexed by $\mathcal{I}(0)=\{1,...,I\}$ and $\mathcal{I}(j)=\{1,...,I(j)\}$,~respectively.
	%where $|\mathcal{I}(0)|\neq|\mathcal{I}(j)|=I_{(j)}$.
	In particular,~$I{(j)}$ and $I$ indicate the total number of users in the $j$-th SBS and MBS, respectively. The MBS and each SBS are equipped with $M_{\text{MC}}$ and $M_{\text{SC}}$ antennas, respectively,~and each RIS is equipped with $N=N_{t,x}N_{t,y}$ passive reflecting elements (a $N_{t,x}\times N_{t,y}$ uniform planar array).~The MC-NOMA scheme is considered for all SBSs, where the whole bandwidth of $B$ [Hertz] is divided into $L$ subcarriers denoted by $\mathcal{L}=\{1,...,L\}$.
	%Note that increasing the number of users on the same subcarrier leads to co-channel interference (CCI), hardware complexity, and processing delay, which can degrade the system performance. Hence,
	It is assumed that each subcarrier is assigned to at most two users in each SBS \cite{Sun}. The set of all µW subcarriers is denoted by $\mathcal{X}=\{1,...,X\}$.
	%In wireless communication networks, the LoS links provide reliable and efficient wireless communications. However, when a blockage occurs, the propagation attenuation of the µW channel links considerably increases. To compensate for the fast attenuation of signals, the MBS is equipped with directional antenna arrays in order to perform beamforming.
	Besides, several UAVs-RISs denoted by $\mathcal{U}=\{1,2,..., U\}$ are deployed to serve the users with the NLoS links,
	%Since each UAV needs to fly at a low altitude to support the users in SBSs,~it is in contrast to the regularity of the UAV to operate at mmW channel.
	and we consider that each UAV-RIS operates at the µW channel for simplicity.
	\subsection{Channel Modeling}
	The channel link between each UAV and each ground user follows the air-to-ground (A2G) communication model \cite{Cai}, and the channel links between the interfering MBS and the SBSs are assumed to follow the Rayleigh flat fading model.
	%\textcolor{blue}{The LoS and NLoS conditions in the A2G (from each UAV to ground users) links are assumed to be encountered randomly.}
	\textcolor{blue}{Define $\mathbf{h}_{{j},i(j)}(t)\in\mathbb{C}^{M_{\text{SC}}}$ as the channel gain from the $j$-th SBS to the $i$-th user in cell $j$ at time instant $t$, which is assumed to experience a quasi-static flat fading.}~The position of the $u$-th UAV-RIS at time instant $t$ is given as $\textbf{q}_u(t)= [x_{u}(t),y_{u}(t),h_u(t)]\in \mathbb{R}^{3}$, where $x_{u}(t)$ and $y_{u}(t)$ are the position of $u$-th UAV-RIS along with $x$ axis and $y$ axis, respectively. $h_{u}(t) \in [h_{\text{min}}(t),h_{\text{max}}(t)]$ is the altitude of the $u$-th UAV-RIS, where $h_{\text{min}}$ and $h_{\text{max}}$ are the minimum and maximum altitudes of each flying UAV-RIS, respectively. The location of the MBS and the $i$-th user in cell $j$ are denoted by $\textbf{q}_{\text{MC}}= [x_{\text{MC}},y_{\text{MC}},h_{\text{MC}}]^T\in \mathbb{R}^{3}$ and $\textbf{q}'_{i(j)}(t)= [x_{i(j)}(t),y_{i(j)}(t),0]^T\in \mathbb{R}^{3}$, respectively. The channel’s power gain between the $u$-th UAV-RIS and the $i(j)$-th user at time instant $t$ is given by
	\	\begin{align}
	&	\mathbf{g}_{u,i(j)}(t)={\sqrt{\eta}{{{{\left\| {\mathbf{{q}}_u(t) - {\mathbf{q}}'_{i(j)}(t)} \right\|}^{-1}}}}} \underbrace{\mathbf{a}^u_{i(j)}(t)}_{\text{AAR}},
	\end{align}where $\eta=(\frac{\lambda}{4\pi})^2$ is a constant
	in which $\lambda$ indicates the wavelength of the center frequency. Besides, $\mathbf{a}^u_{i(j)}(t)\in {\mathbb{C}^{N \times 1}}$ is the antenna array response (AAR) between the $u$-th UAV-RIS and ${i(j)}$-th user at time instant $t$. In particular, the AAR vector is represented by
	\begin{align}
	{\mathbf{a}^u_{i(j)}}(t) &= \left(1,{e^{ - j\frac{{2\pi d^{u}_{i(j)}(t)}}{\lambda }\sin {\theta^u _{i(j)}}(t)\cos {\varphi^u _{i(j)}}(t)}},...,{e^{ - j\frac{{2\pi d^{u}_{i(j)}(t)}}{\lambda }\sin {\theta^u _{i(j)}}(t)({N_x} - 1)\cos {\varphi^u _{i(j)}}(t)}}\right)\nonumber\\ &\otimes \left(1,{e^{ - j\frac{{2\pi d^{u}_{i(j)}(t)}}{\lambda }\sin {\theta^u _{i(j)}}(t)\cos {\varphi^u _{i(j)}}(t)}},...,{e^{ - j\frac{{2\pi d^{u}_{i(j)}(t)}}{\lambda }\sin {\theta^u _{i(j)}}(t)({N_y} - 1)\cos {\varphi^u _{i(j)}}(t)}}\right),
	\end{align}where $d^{u}_{i(j)}(t)$ is the distance between each reflecting element of the $u$-th UAV-RIS and each user at time instance $t$. In addition, $\theta^{u} _{i(j)}(t)$ and $\varphi^{u} _{i(j)}(t)$ are the vertical and horizontal angle of departure (AoD) between the $u$-th UAV and ${i(j)}$-th user at time instance $t$, respectively.~Consequently, the AoDs are given by
	\begin{align}
	&\theta^{u}_{i(j)}(t)=\arcsin\bigg(\frac{h_u(t)}{\|\textbf{{q}}_u(t) - {\mathbf{q}}'_{i(j)}(t)\|}\bigg),\\
	&\varphi^{u} _{i(j)}(t)=\arccos\bigg(\frac{{y_u(t) - {{y_{i(j)}}(t)}}}{{\sqrt {{{(x_u(t) - {{x_{i(j)}}}(t))}^2} + {{(y_u(t) - {{y_{i(j)}}(t)})}^2}} }}\bigg).
	\end{align}	
	%	However, $\mathbf{a}_{i(j)}(t)$ is nonlinear functions, which make the design of robust resource allocation difficult. To deal with this issue, the fRISt order Taylor series expansion is employed to approximate $\mathbf{a}_k(t)$ as follows
	%	\begin{align}
	%	&{{\bf{a}}_k}(t) \approx {{\bf{\tilde a}}_k}(t) + {\left. {{{(\nabla {{\bf{a}}_k}(t))}_{{\theta _k}(t)}}} \right|_{\scriptstyle{\theta _k}(t) = {{\hat \theta }_k}(t)\atop\scriptstyle{\varphi _k}(t) = {{\hat \varphi }_k}(t)}}\Delta {\theta _k}(t) + {\left. {{{(\nabla {{\bf{a}}_k}(t))}_{{\varphi _k}(t)}}} \right|_{\scriptstyle{\theta _k}(t) = {{\hat \theta }_k}(t)\atop\scriptstyle{\varphi _k}(t) = {{\hat \varphi }_k}(t)}}\Delta {\varphi _k}(t),
	%	\end{align}
	%	where $\hat{\mathbf{a}}_k(t)$ expresses estimation of the AAR for $k$-th CU receiver which is given by
	%	\begin{align}
	%	&\tilde{\mathbf{a}}_k(t)={\mathbf{a}}({\theta _k}(t),{\varphi _k}(t))|_{\scriptstyle{\theta _k}(t) = {{\hat \theta }_k}(t)\atop\scriptstyle{\varphi _k}(t) = {{\hat \varphi }_k}(t)}.
	%	\end{align}
	\textcolor{blue}{The channel link between the $u$-th UAV-RIS and the MBS at time instant $t$ is represented as $\mathbf{G}_{0,u}(t) \in {\mathbb{C}^{N\times M_\text{MC}}},$ which is assumed to follow LoS model similar to $\mathbf{g}_{ui(j)}(t)$.} Note that due to the possible movement of each user and UAVs-RISs as well as the blockage effect, obtaining the prefect CSI 
	%of the RISs-MUEs links 
	is quite challenging \cite{Bennis1}\footnote{\textcolor{blue}{In general, with the existence of RF chains at the IRS, traditional channel estimation techniques can be used to estimate the channel links of the AP-IRS and IRS-user [15]. With the absence of RF chains, uplink pilots in conjunction with IRS reflection patterns can be designed to estimate the corresponding channel links [26].}}. Thus, the real-time measurement is required to determine the value of channel gain.~However, it is impractical to move each UAV-RIS to all possible locations to acquire perfect CSI since such a sweeping search will consume significant power and time. To address the challenge of the CSI estimation, a learning-based approach can be applied to enable efficient deployment of the UAVs-RISs.
	\subsection{Transmission Scheme}
	In this section, we provide a detailed analysis of the transmission strategy for different users, i.e., MUEs and SUEs.~Without loss of generality, it is assumed that the transmit signal, i.e., $x_{j}$, has zero mean and unit variance.~\textcolor{blue}{`Let $z^{[l]}_{i(j),i'(j)}(t)$ be the subcarrier assignment indicator for each SBS at time instant $t$ in which user $i$ and user $i'$ belong to the same cell ($j$) defined as
	\begin{equation*}
	z^{[l]}_{i(j),i'(j)}(t)=
	\begin{cases}
	1, & \text{if user $i$ and user $i'$ are multiplexed}\nonumber\\& \text{over subcarrier $l \in \mathcal{L}$ at time instant $t$,} \\
	0, & \text{otherwise},
	\end{cases}
	\end{equation*}}
	\begin{equation*}
	\rho^{[x]}_{i^{''}(j)}(t)=
	\begin{cases}
	1, & \text{if user $i^{''}$ is asssigned to subcarrier} \nonumber\\&\text{$x \in \mathcal{X}$ at time instant $t$}, \\
	0, & \text{otherwise},
	\end{cases}
	\end{equation*}
	respectively. By assuming that each subcarrier includes two PD-NOMA users, $i$ and $i'$, the received signal at user $i(j)$ over subcarrier $l$ for the mmW channel at time instant $t$ is given by
	\begin{figure}
		\centering
		\includegraphics[width=4.5in] {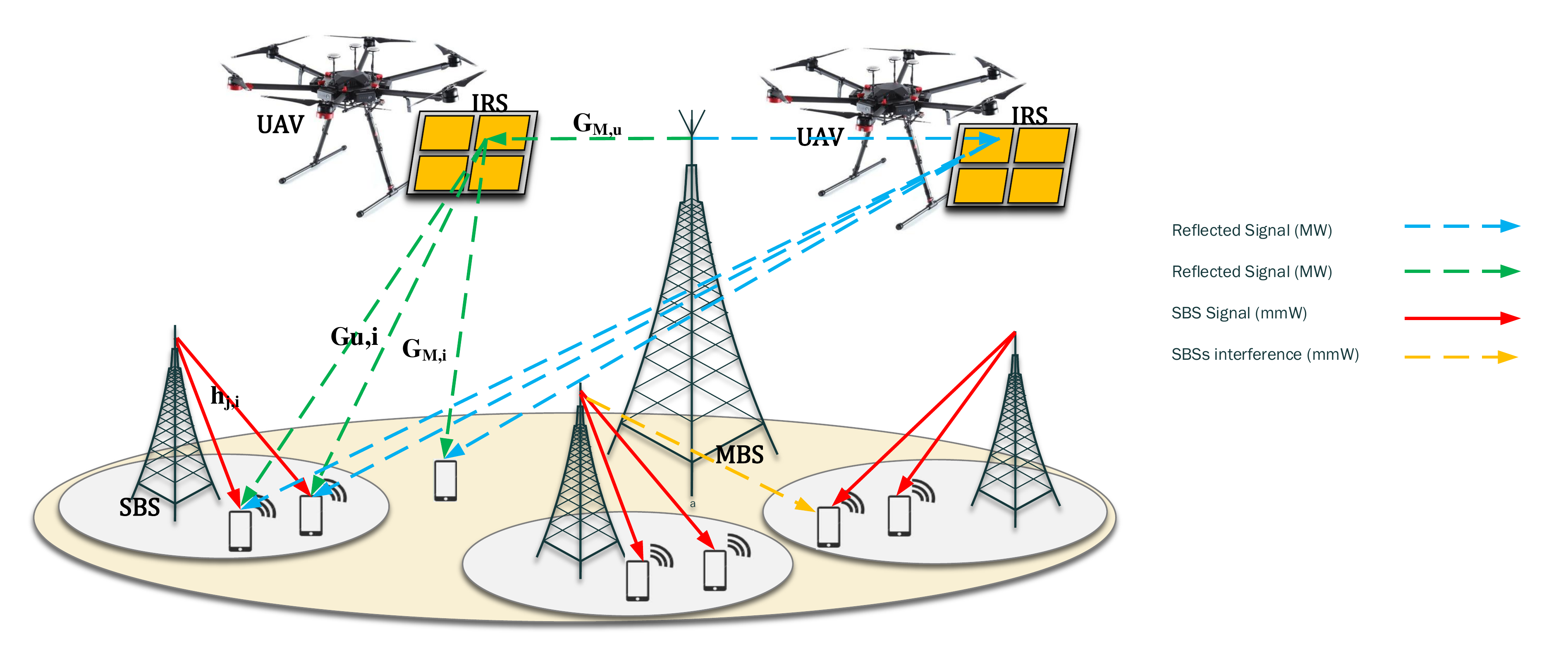}
		\caption{\small \textcolor{blue}{Deployment of the UAVs-RISs in a HetNet where SBSs and MBS apply the MC-PD-NOMA and OFDM schemes, respectively.}}
	\end{figure}
	\begin{align}
	y_{i(j)}^{[l],\text{mmW}}(t)&=	z^{[l]}_{i(j),i'(j)}(t)\Big(\underbrace{\mathbf{h}^{[l]H}_{j,i(j)}(t)\mathbf{w}^{[l]}_{j,i(j)}(t)}_{\text{Desired signal}} +\underbrace{\mathbf{h}^{[l]H}_{j,i(j)}(t)\mathbf{w}^{[l]}_{j,i'(j)}(t)}_{\text{PD-NOMA interference}}\Big)+\nonumber\\&\underbrace{\sum_{j'\neq j}\sum_{k(j')}	z^{[l]}_{i(j),k(j')}(t)\mathbf{h}^{[l]H}_{j',i(j)}(t)\mathbf{w}^{[l]}_{j',k(j')}(t)}_{\text{Interference of other cells}}+{\xi^{[l],\text{mmW}}},
	\end{align}where $\mathbf{w}^{[l]}_{j,i(j)}(t)$ denotes the beamforming vector of the $j$-th SBS to user $i$ in cell $j$, $\mathbf{h}^{[l]H}_{j,i(j)}$ indicates the channel link between user $i$ in cell $j$ and the $j$-th SBS, $\mathbf{w}^{[l]}_{j',k(j')}(t)$ is the beamforming vector of the $j'$-th SBS to user $k$ in cell $j'$ over subcarier $l$,~and finally 
	%More specifically, the first term is the desired signal, the second term is due to the PD-NOMA interference in the same cell.~The third one is the interference of other cells or other SBSs, and 
	$\xi^{[l],\text{mmW}}\sim \mathcal{N}(0,\,\sigma^{2})\,$ is the receiver noise on mmW channel over subcarrier $l$, which is assumed to be the same for all the SUEs.
	%Note that RISs are mounted under the UAVs \cite{Alfattani}, so the UAVs-RISs cannot have a severe effect on each other. Besides, the locations of the users and UAVs-RISs are sufficiently far apart, such that the interference of the other UAVs-RISs at a specific user is negligible as a result of the high path-loss \cite{Rui5}. Moreover, the reflected signals from other UAVs-RISs to the intended UAV-RIS have a little impact on the performance, since each UAV-RIS steers its phase shifts according to the intended user.
	
\textcolor{blue}{Based on the PD-NOMA scheme, each user attempts to adopt SIC to decode its own signal by treating the other signals as noise. There is a condition on the SINR for the $i$-th user with a better DL channel condition to detect the $i'$-th user's data with a worse one over subcarrier $l$ in the SBS. In fact, on subcarrier $l$, the following condition should be held to ensure that user $i$ can decode and remove the interference from user $i'$ successfully:}
\textcolor{blue}{\begin{align}\label{noma}
	&J^{[l]}_{i(j),i'(j)}(\mathbf{w}(t))=\log_2\bigg(1+\frac{|\mathbf{h}^{[l]H}_{j,i'(j)}(t)\mathbf{w}^{[l]}_{j,i'(j)}(t)|^{2}}{\sum_{j'\neq j\backslash \{ 0 \}}\sum_{k(j')}|\mathbf{h}^{[l]H}_{j',i'(j)}(t)\mathbf{w}^{[l]}_{k(j')}(t)|^{2}+|\mathbf{h}^{[l]H}_{j,i'(j)}(t)\mathbf{w}^{[l]}_{j,i(j)}(t)|^{2}+\sigma^{2}}\bigg)\nonumber\\&\qquad\qquad\qquad-\log_2\bigg(1+\frac{|\mathbf{h}^{[l]H}_{j,i(j)}(t)\mathbf{w}^{[l]}_{j,i'(j)}(t)|^{2}}{\sum_{j'\neq j\backslash \{ 0 \}}\sum_{k(j')}|\mathbf{h}^{[l]H}_{j',i(j)}(t)\mathbf{w}^{[l]}_{k(j')}(t)|^{2}+|\mathbf{h}^{[l]H}_{j,i(j)}(t)\mathbf{w}^{[l]}_{j,i(j)}(t)|^{2}+\sigma^{2}}\bigg)\leq 0.
	\end{align}}
	The received signal over the µW channel can be written as \footnote{Note that RISs are mounted under the UAVs \cite{Alfattani}, so the UAVs-RISs cannot have a severe effect on each other. Besides, it is assumed that the locations of the users and UAVs-RISs are sufficiently far apart, such that the interference of the other UAVs-RISs at a specific user is negligible as a result of the high path-loss \cite{Rui5}.} %Moreover, the reflected signals from other UAVs-RISs to the intended UAV-RIS have a little impact on the performance, since each UAV-RIS steers its phase shifts according to the intended user.}
	\begin{equation}
	y_{i^{''}(j)}^{[x],\text{µW}}(t)=\rho^{[x]}_{i^{''}(j)}(t)\left\{\sum_{u \in \mathcal{U}}\mathbf{g}_{u,i^{''}(j)}^{[x],H}(t) \mathbf{\Theta}_u(t)\mathbf{G}^{[x]}_{0,u}(t) \mathbf{w}^{[x]}_{0,i^{''}(j)}(t)\right\}+\xi^{{[x]},\text{µW}},
	\end{equation}where $\xi^{^{[x]},\text{µW}}\sim \mathcal{N}(0,\,\delta^{2})\,$ is the receiver noise over subcarrier $x$, which is assumed to have the same distribution for all users.~More specifically, $\mathbf{w}^{[x]}_{0,i^{''}(0)}$ is the beamforming vector for user $i^{''}$ in macro-cell over subcarrier $x$;~$\mathbf{G}^{[x]}_{0,u}(t)\in \mathbb{C}^{N\times M}$ and $\mathbf{g}^{[x]}_{u,i^{''}(0)}(t) \in \mathbb{C}^{N }$ denote the channel responses from the MBS to the $u$-th UAV-RIS and each UAV-RIS to user $i^{''}$ in macro-cell over subcarrier $x$,~respectively.~\textcolor{blue}{Also,~$\mathbf{\Theta}_u(t)=\text{diag}(a_{1}(t)e^{j\theta_{1}(t)},..., a_{N}(t)e^{j\theta_{N}(t)})$ represents the phase shift matrix of the $u$-th UAV-RIS where $\theta_{n}(t) \in [0,2\pi]$ and $a_{n} (t)\in (0,1]$ denotes the phase and amplitude  of each reflecting element, respectively. In particular, we consider discrete phase and amplitude for each reflecting element by dividing $\theta_{n}(t)$ and $a_{n}(t)$ into $\frac{2\pi}{100}$ and $\frac{1}{0.1}$, respectively.}
	%when $\mathbf{h}^{[l]}_{j,i(j)}\geq\mathbf{h}^{[l]}_{j,i'(j)}$ and
	%	where $I_{\text{CCI}}(t)=\sum_{j'\neq j}\sum_{i(j')}|\mathbf{h}^{[l]H}_{j',i(j)}(t)\mathbf{w}^{[l]}_{i(j')}(t)|^{2}$ is due to the interference of other SBSs.
	%Furthermore, we introduce  $\mathbf{W}\in\mathbb{C}^{LI\times M_{sc}}$ and $\mathbf{s}\in\mathbb{C}^{LI\times 1}$ for simplicity.
	Note that we remove the unknown parameter $\mathbf{\Theta}$ from the subcarrier allocation policy and assumed that RIS reflects each received signal over all subcarriers. In addition, it is assumed that the amplitudes of each RIS are equal to one (i.e., $a_{n} (t)=1$, $\forall n$).~Consequently, the data rate for users $i(j)$ and $i'(j)$ over the subcarrier $l$ for the mmW channel can be expressed as
\begin{align}
	&R^{[l],\text{mmW}}_{i(j)}(t)= z^{[l]}_{i(j),i'(j)}(t)\log_2\bigg(1+\frac{|\mathbf{h}^{[l]H}_{j,i(j)}(t)\mathbf{w}^{[l]}_{j,i(j)}(t)|^{2}}{\sum_{j'\neq j\backslash \{ 0 \}}\sum_{i''(j')}|\mathbf{h}^{[l]H}_{j',i(j)}(t)\mathbf{w}^{[l]}_{j',i''(j')}(t)|^2+\sigma^{2}}\bigg),\label{rate22}
\end{align}
{\small \begin{align}
	&R^{[l],\text{mmW}}_{i'(j)}(t)= z^{[l]}_{i(j),i'(j)}(t)\log_2\bigg(1+\frac{|\mathbf{h}^{[l]H}_{j,i'(j)}(t)\mathbf{w}^{[l]}_{j,i'(j)}(t)|^{2}}{\sum_{j'\neq j\backslash \{ 0 \}}\sum_{i''(j')}|\mathbf{h}^{[l]H}_{j',i'(j)}(t)\mathbf{w}^{[l]}_{j',i''(j')}(t)|^2+|\mathbf{h}^{[l]H}_{j,i'(j)}(t)\mathbf{w}^{[l]}_{j,i(j)}(t)|^{2}+\sigma^{2}}\bigg),
	\end{align}}respectively.
	%In this work, we suppose that the MBS can only serve one user in each time slot.~To this end, we define $d_{i(0)}(t)\in \{0,1\}, \forall i(0), t$,~where $d_{i(0)}(t)=1$ denotes that each MBS sends data to MUE in time slot $t$, otherwise it would be zero.~We further assume that each MBS always communicate to the MUE \textcolor{blue}{with maximum} data rate, i.e., \textcolor{blue}{$d_{i(0)}(t)=1,~i(0)=\text{argmax}_{i(0)} (\mathbf{g}^{H}_{0,i(0)}(t))$}.~
	As a result, the sum rate 
	%of users $i$ and $i'$ in cell $j$ over the subcarrier $l$
	can be written as $R^{[l],\text{mmW}}_{i(j),i'(j)}(t)=R^{[l],\text{mmW}}_{i(j)}(t)+R^{[l],\text{mmW}}_{i'(j)}(t)$.~Consequently,~the data rate for user $i^{''}(j)$ over the subcarrier $x$ for the µW channel can be written as
	%\textcolor{blue}{	\begin{align}\label{rate2}
	%	R^{\text{µW}}_{i(j)}(t)=&d_{i(0)}(t)\log_2\bigg(1+\frac{\mathbf{g}_{u,i(0)}^H\mathbf{w}_{0,i(0)}(t)+|\sum_{j\in \mathcal{J}}\sum_{u=1}^U \mathbf{g}_{u,i(j)}^H(t) \mathbf{\Theta}_u(t)\mathbf{G}_{0,u}(t) \mathbf{w}_{0,i(j)}(t)|^{2}}{\delta^{2}}\bigg).
	%	\end{align}}
	\begin{align}\label{rate2}
	R^{[x],\text{µW}}_{i^{''}(j)}(t)=&\log_2\bigg(1+\frac{\sum_{u\in \mathcal{U}}| \rho^{[x]}_{i^{''}(j)}(t)\mathbf{g}_{u,i^{''}(j)}^{[x],H}(t) \mathbf{\Theta}_u(t)\mathbf{G}^{[x]}_{0,u}(t) \mathbf{w}^{[x]}_{0,i^{''}(j)}(t)|^{2}}{\delta^{2}}\bigg).
	\end{align}
	Furthermore, we suppose that each UAV-RIS can track the location of each user to establish a DL µW channel link.
	% with the aid of the RIS.
	%In the following, we fRISt define the start location and end location of the UAV by $\mathbf{q}_s=[x_{s,u},y_{s,u},h_{s,u}]$ and $\mathbf{q}_e=[x_{e,u},y_{e,u},h_{e,u}]$, respectively, for the coherence time interval of wireless communication given by $0\leq t\leq T$. Consequently, the minimum distance from the start to the end location in the coherence time interval $T$ is given by $d_\text{min}=\|\mathbf{q}_e-\mathbf{q}_s\|$. By denoting the UAV’s maximum velocity as $V$, we presume that $V\geq \frac{d_\text{min}}{T}$ to fined at least a feasible path.
	%The channel $\mathbf{g}_{u,i(j)}$ is assumed to be fixed in one coherence time $ t$.
	However, the UAVs' trajectories change continuously over time, which makes the optimization problem and derivations intractable due to the integral operations. To overcome this, we discretize the total flying time interval, denoted by $T$, into $K$ time slots with equal spacing, i.e., $T = K\tau$, where $K \in \mathbb{N}$ and $\tau$ is assumed to be sufficiently small such that the each UAV’s position remains relatively constant in each time slot. As a result, the resource allocation policy can be designed for each time slot $\tau$.
	% In other words, the movement of the RIS makes the beamforming process much more difficult.~To facilitate the system design,~we assume that communication occurs during $\tau$, where the position of each UAV-RIS is static. 
	Accordingly, the UAV’s mobility constraints can be expressed as
	\begin{align}
	%&\|\mathbf{q}(1)-\mathbf{q}_s\|^2\leq (Vt)^2,\\
	\mathbf{q}_u(t+1)=\mathbf{q}_u(t)+ \mathbf{v}_{u}(t)\tau,~\forall t \in \{1,...,K\},
	%&\|\mathbf{q}_e-\mathbf{q}(T)\|^2\leq (Vt)^2,
	\end{align}
	where $\mathbf{v}_u(t)=[v_{u,x}(t),v_{u,y}(t),v_{u,z}(t)]^T\in\mathbb{R}^{3\times 1}$ is the flight velocity of each UAV-RIS.
	\subsection{Power Consumption Model}
\textcolor{blue}{The total transmit power of the considered HetNet system, which involves the transmit power of the SBSs and MBS is given by}
\textcolor{blue}{\begin{align}
	P_\text{total}(t)&=\frac{1}{\Xi_\text{SBS}}~\underbrace{\sum_{j=1}^{J}\sum_{l=1}^{L}\sum_{i=1}^{I(j)}\sum_{i'=1}^{I(j)} z^{[l]}_{i(j),i'(j)}(t)\left\{\|\mathbf{w}^{[l]}_{j,i(j)}(t)\|^{2}+\|\mathbf{w}^{[l]}_{j,i'(j)}(t)\|^{2}\right\}}_{\text{SBS}}\nonumber\\&+\frac{1}{\Xi_\text{MBS}}\underbrace{\sum_{j=1}^{J}\sum_{x=1}^{X}\sum_{i^{''}=1}^{I(j)} \rho^{[x]}_{i^{''}(j)}(t)\|\mathbf{w}^{[x]}_{0,i^{''}(j)}(t)\|^{2}}_{\text{MBS}},
	\end{align}}		 \textcolor{blue}{where $0<\Xi_\text{SBS}<1$ and $0<\Xi_\text{MBS}<1$ are the drain efficiencies of the power amplifiers.}		

	\section{Problem Formulation}\label{ProblemForm}
	In this section, we formulate an optimization problem to
	jointly optimize the active beamformers at the MBS/SBSs, the phase shift matrix at each RIS, UAVs' trajectories/velocities, and subcarrier allocations for µW/mmW,~which aims to minimize the total transmit power of the system.~Mathematically speaking, the optimization problem can be formulated as
	\begin{align}\label{M_U}
\text{P1}:~& \underset{\mathbf{z},\boldsymbol{\rho},\mathbf{q}_u,\mathbf{w},\mathbf{\Theta}_u,\mathbf{v}} {\text{minimize}}\:\frac{1}{T}\sum_{t=1}^{T} P_\text{total}(t)\nonumber \\
&\text{s.t.}~	
	C_{1}:~\sum_{l=1}^{L}\sum_{i=1}^{I(j)}\sum_{i'=1}^{I(j)} z^{[l]}_{i(j),i'(j)}(t)\left\{\|\mathbf{w}^{[l]}_{j,i(j)}(t)\|^{2}+\|\mathbf{w}^{[l]}_{j,i'(j)}(t)\|^{2}\right\}\leq P^{j}_{\max},\:\forall t, \forall j \in \mathcal{J}, \:\text{if}:\: j\neq 0,\nonumber\\
	&\quad~ C_{2}:~\sum_{x=1}^{X}\sum_{i^{''}=1}^{I(j)} \rho^{[x]}_{i^{''}(j)}(t)\|\mathbf{w}^{[x]}_{0,i^{''}(j)}(t)\|^{2}\leq P^{0}_{\max},~\forall t, \:\text{if}:\:j= 0,\nonumber\\
	&\quad~ C_{3}:~\sum_{l=1}^{L}R^{[l],\text{mmW}}_{i(j),i'(j)}(t)+\sum_{x=1}^{X}R^{[x],\text{µW}}_{i^{"}(j)}(t)\geq R^{i}_{\min},~\forall i,i'\in \mathcal{I}(j), \forall i^{''}\in \mathcal{I}(0), \forall t,	\nonumber\\
	&\quad~C_{4}:~\sum_{l=1}^{L}R^{[l],\text{mmW}}_{i(j),i'(j)}(t)\geq \tilde{R}^{i}_{\min},~\forall i,i'\in \mathcal{I}(j), \forall t, \forall j \in \mathcal{J},\:\text{if}:\: j\neq 0,\nonumber\\
	&\quad~C_{5}:~0\leq\theta^{u}_{n}(t)\leq 2\pi,~\forall u \in \mathcal{U},\forall t,
	\quad~C_{6}:~h_{\text{min}}(t)\leq h_{u}(t) \leq h_{\text{max}}(t),~\forall u \in \mathcal{U},\forall t,	\nonumber\\
	&\quad~C_{7}:~\mathbf{q}_u(t+1)=\mathbf{q}_u(t)+ \mathbf{v}_{u}(t)\tau,~\forall u \in \mathcal{U}, \forall t,	\nonumber\\
	&\quad~C_{8}:~\|\mathbf{v}_u(t+1)-\mathbf{v}_u(t)\|\leq a_\text{u} \tau,~\forall u \in \mathcal{U}, \forall t,\nonumber\\
	&\quad~C_{9}:~\|\mathbf{v}_u(t)\|\leq V_\text{u,max},~\forall u \in \mathcal{U}, \forall t \nonumber,\\
	&\quad~C_{10}:~\sum_{i=1}^{I(j)}\sum_{i'=1}^{I(j)}z^{[l]}_{i(j),i'(j)}(t)\leq 1,~\forall l\in \mathcal{L}, \forall t, \forall j \in \mathcal{J},\:\text{if}:\: j\neq 0, \nonumber\\
	&\quad~C_{11}:~z^{[l]}_{i(j),i'(j)}(t)J^{[l]}_{i(j),i'(j)}(\mathbf{w}(t))\leq 0,~\forall l\in\mathcal{L},\:\forall i,i'\in \mathcal{I}(j), \forall t, \forall j \in \mathcal{J}, \:\text{if}:\:j\neq 0,\nonumber\\
	&\quad~C_{12}:~z^{[l]}_{i(j),i'(j)}(t)\in\{0,1\},~\forall l\in\mathcal{L},\:\forall i, i'\in \mathcal{I}(j), \forall t, \forall j \in \mathcal{J},~ \:\text{if}:\:j\neq 0,\nonumber\\
	&\quad~C_{13}:~\sum_{i^{''}=1}^{I(j)}\rho^{[x]}_{i^{''}(j)}(t)\leq 1,~\forall x\in \mathcal{X}, \forall t, \:\text{if}:\: j=0,\nonumber\\
	&\quad~C_{14}:~\rho^{[x]}_{i^{''}(j)}(t)\in\{0,1\},~\forall i^{''}\in \mathcal{I}(0), \forall x\in \mathcal{X}, \forall t, \:\text{if}:\:j=0,
	\end{align}where,~$\mathbf{z}$ and $\boldsymbol{\rho}$ are the collection of subcarrier allocations for the mmW and µW, respectively. $\mathbf{q}_u$ denotes the trajectory of the $u$-th UAV, $\mathbf{w}$ is the collection of beamforming vectors, $\mathbf{\Theta}_u$ denotes the phase shifts,~and finally $\mathbf{v}$ indicates the vector of velocities.~$C_{1}$ and $C_{2}$ limit the total transmit power of each SBS and MBS, respectively.~$C_{3}$ and $C_{4}$ guarantee the minimum quality of service (QoS) for each user in macro-cell and small-cells, respectively.~$C_{5}$ is the phase shift constraint of the $u$-th UAV-RIS's reflecting elements.~$C_{6}$ limits the minimum and maximum altitudes of each flying UAV, respectively.~$C_{7}$ is the mobility constraint of each UAV-RIS.~$C_{8}$ and $C_{9}$  denote the acceleration and maximum flight velocity, respectively. $C_{10}$ ensures that each subcarrier in each SBS is allocated to at most two SUEs\footnote{\textcolor{blue}{It is worth mentioning that our work can be extended to scenarios with more than two users by revising constraint ${C}_{10}$ and SIC decoding order.}}.~$C_{11}$ guarantees successful SIC at SUE $i$ if $z^{[l]}_{i(j),i'(j)}(t)=1$.~$C_{12}$ indicates the subcarrier assignment for each SUE which is a binary variable constraint.~$C_{13}$ ensures that each subcarrier in MBS is merely assigned to one MUE in time slot $t$.~Finally, $C_{14}$ indicates the assignment for each MUE in each time slot which is a binary value.
	\begin{figure}[t]
		\centering
		\includegraphics[width=3in]{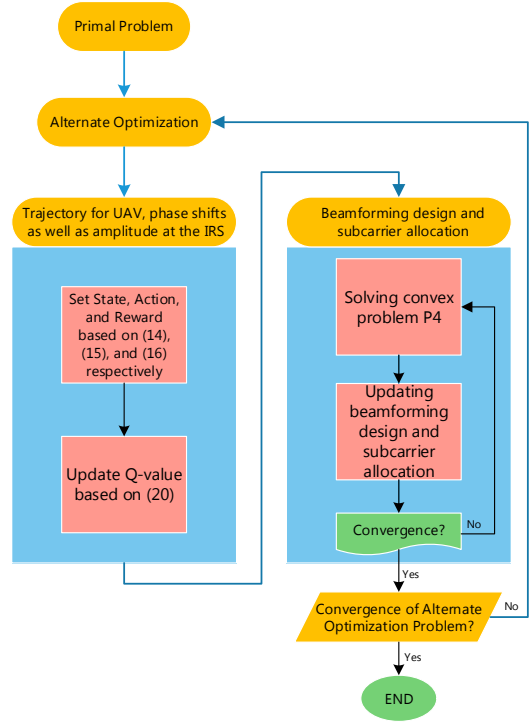}
		\caption{\small The hierarchically of the proposed solution.}
			\vspace{-5mm}
	\end{figure}
		\vspace{-5mm}
	\section{Proposed Algorithm}
	It can be seen that (P1) is a mixed integer programming problem, which is difficult to solve. However, (P1) can be divided into two sub-problems, namely: (i) UAV's trajectory ($\mathbf{q}_u$), RIS's phase shifts ($\mathbf{\Theta}_u$), and MBS subcarrier allocation ($\boldsymbol{\rho}$) sub-problem; (ii) joint beamforming ($\mathbf{w}$) design and SBSs subcarrier allocation ($\mathbf{z}$)
	sub-problem.~\textcolor{blue}{It should be noted that optimizing all the variables via learning methods is difficult as the size of the problem increases significantly. Besides, the mini-batch size is limited, and the RL algorithm can be used for discrete variables. Thus, for the beamforming solution, we obtain a locally optimal solution via the convex optimization method to avoid inaccurate answers.}~Therefore, a two-stage decision making procedure is considered to deal with the original complicated problem. For the first stage, by using the Deep Q-learning technique, the UAV-RISs take one of the available actions to obtain their trajectories/velocities, phase shifts, and subcarrier allocation. Upon any change in the position of each UAV, phase shift of each RIS, and SBSs subcarrier allocation in the second stage, a new optimization problem is solved to determine passive beamforming and MBS subcarriers allocation.
	%The corresponding throughput contributes to the reward received from taking the action in the first stage.
	In fact, we adopt an alternative optimization (AO) method where the original problem is divided into two subproblems.~The hierarchically of the solution methodology is given in Fig. 2. In particular, we define the agents, states, actions, and reward associated with the Q-learning framework in the following:
	
	\textbf{Agent} $u$: UAV-RIS $u$, $\forall u
	\in \mathcal{U}$.
	
	\textbf{State} $\mathcal{S}_{t}$: The position of the $u$-th UAV-RIS at time slot $t$, i.e., $\textbf{q}_u(t)= [x_{u}(t),y_{u}(t),h_{u}(t)]$.
	%	 Note that if we consider a continuous position for each UAV, it is hard to address the problem. To deal with this issue, we discretize the given area by considering a grid with a restricted number of squares such that each square's center describes that state. In this way, the number of possible states can be reduced. For instance,
	We consider a continuous rectangular region with the dimension of $x_\text{min} \leq x \leq x_\text{max}$ and $y_\text{min} \leq y \leq y_\text{max}$, then we transform it into $K^3$ states such that both intervals $[x_\text{min},x_\text{max}]$ and $[y_\text{min},y_\text{max}]$ are divided into $K$ slots. The coordination of the square's center can be represented by $x_{k_{1}} = x_\text{min}+ \frac{(x_\text{max}-x_\text{min})}{K}(k_{1} -1)$, $y_{k_{2}} = y_\text{min} + \frac{(y_\text{max}-y_\text{min})}{K}(k_{2} -1)$, and $h_{k_{3}} = h_\text{min} + \frac{(h_\text{max}-h_\text{min})}{K}(k_{3} -1)$ at the $k_1$-th slot in the $x$ axis, at the $k_2$-th slot in the $y$ axis, at the $k_3$-th slot in the $z$ axis, respectively.~Also, the channel states of all users as well as UAVs-RISs are the states of the model. Consequently, the state space is defined as
	\begin{equation}
	\mathcal{S}_{t}=\{\textbf{q}_u(t),\mathbf{G}_{0,u}(t),\mathbf{h}_{{j},i(j)}(t),\mathbf{g}^{[x]}_{u,i^{''}(0)}(t)\}.
	\end{equation}
	
	\textbf{Action} $\mathcal{A}_{u}(t)$: Possible actions for the movement of each UAV at each state are left, right, forward, backward, up, and down as well as its speed. Besides, possible action for each RIS is adjusting phase shifts.~We also define the feasible action for each UAV as $\mathcal{q}_u$,~which includes left, right, forward, backward, up, and down as well as its speed. Hence, the action vector is given by
	\begin{equation}
	\mathcal{A}_{(t)}=\{\mathcal{q}_u,\theta_{u1},...,\theta_{uN},\rho^{[x]}_{i^{''}(j)},\mathbf{v}_{u}\}.
	\end{equation}
	
	\textbf{Reward} $\mathcal{R}_{(t)}$: The reward functions are commonly related
	to the objective of the problem. The reward function is defined as
	%	\begin{equation}
	%	r(t)=F_1(t)-F_2(t),
	%	\end{equation}
	%	where $F_1(t)$ denotes the sum-rate of the users of the network and $F_2(t)$ is the negative reward which acts as a punishment.~The function $F_1(t)$ is given by:
	%	\begin{align}
	%	F_1(t)\triangleq\sum_{j}\sum_{i}\sum_{i'}\sum_{l}z^{*[l]}_{i(j),i'(j)}(t)R^{[l],\text{mmW}}_{i'(j)}(\mathbf{w}^{*}(t))+R^{\text{µW}}_{i(j)}(\mathbf{w}_{0}^{*}(t)),
	%	\end{align}
	%	where $s^{*[l]}_{i(j),i'(j)}(t)$ and $\mathbf{w}^{*}(t)$ are the optimal subcarrier assignment and beamforming design at time slot $t$,~which will be addressed later in subsection $C$. $F_2(t)$ is the minimum data rate requirement which is given by
	%	\begin{equation}
	%	F_2(t)\triangleq \sum_{l=1}^{L}R^{[l],\text{mmW}}_{i(j)}(t)+R^{\text{µW}}_{i(j)}(t)\geq R^{i}_{\min},\:\forall i\in \mathcal{I}, \forall j\in \mathcal{J}, \forall t.
	%	\end{equation}
	%	Now, we can restate the reward function as follows
	\begin{align}
	\mathcal{R}{(t)} =
	\begin{cases}
	\frac{1}{P_\text{total}(t)},& \text{if all constraints are satisfied},\\
	{0},& \text{otherwise}.
	\end{cases}
	\end{align}
	The above-proposed reward function increases with decreasing power consumption.~Therefor it may finally achieve the minimum of the energy consumption of the networks.
	%    where $D_u(t)$ denotes the distance between $u$-th UAV and its final
	%    position at time instant $t$ which is define as $D_u(t)=\|\mathbf{q}_u(t)-\mathbf{q}_{u,F}(t) \|^2$, $\forall t,u$, where $\textbf{q}_{u,F}(t)= [x_{u,F}(t),y_{u,F}(t),h_{u,F}(t)]$. Furthermore, to avoid the collision of UAVs, activation function $F^{(u)}_3(t)$ is introduced as
	%    \begin{align}
	%        F^{(u)}_3(t)= \left\{ \begin{array}{l}1{\rm{     }}{{\quad\quad\rm{D}}_{u,u'}}(t) < {D_{\text{min} }},\forall u \ne u',\\0{\rm{   \quad\quad \text{Otherwise}}}{\rm{.}}\end{array} \right.
	%    \end{align}

\subsection{Multi-Agent Q-learning}
In this subsection, we first define preliminary definitions for the learning method, and then we employ dueling DQN to solve the optimization problem. Here, we model the joint optimization as a Markov decision process (MDP) $(\mathcal{S},\mathcal{A},\mathcal{R},\mathcal{P}_{ss'})$, in which $\mathcal{S}$ is the set of environment states, $\{\mathcal{A}_{1},...,\mathcal{A}_{U}\}$ is a discrete set of possible actions of UAV, $\mathcal{R}$ is the reward function, and $\mathcal{P}_{ss'}$ is the state transition probability from state $s$ to state $s'$.~The optimal policy for maximizing the long-term reward for multi-agent by using reinforcement learning in an unknown stochastic environment can be obtained as $\pi^{*}_{u}: \mathcal{S}\rightarrow\mathcal{A}_{u}$.~For the UAVs, the optimal policy $\pi^{*}_{u}$ to maximize its value state function at each state is given by
\begin{align}
V(s,\pi_{u},\pi_{-u})=\mathbb{E}\Big[\sum\limits_{t=0}^{T-1}\gamma^t \mathcal{R}\big(s(t),\pi_{u}(t),\pi_{-u}(t)\big)|s(0)=s\Big],
\end{align}
where $\pi_{-u}$ indicates the vector of other $N-1$ agents, i.e., $\pi_{-u}=(\pi_{1},...,
\pi_{u-1},\pi_{u+1},...,\pi_{U})$.~By considering the Markov property, the value function can be stated as
\begin{align}
V(s,\pi_{u},\pi_{-u})=&\mathbb{E}\Big[ \mathcal{R}\big(s,\pi_{u}(t),\pi_{-u}(t)\big)\Big]+\gamma\sum_{s'\in \mathcal{S}} P_{ss'}(\pi_{u},\pi_{-u})V(s',\pi_{u},\pi_{-u}),
\end{align}
where $\gamma$ is the discount rate.~Consequently, based on the Bellman optimality equation, the optimal value for $Q^{*}_{u}(s,a_{u})$ can be achieved as
\begin{align}
Q^{*}_{u}(s,a_{u})=\mathbb{E}\Big[ \mathcal{R}\big(s,\pi_{u}(t),\pi^{*}_{-u}(t)\big)\Big]+\gamma\sum_{s'\in \mathcal{S}} P_{ss'}(a_{u},\pi^{*}_{-u})\max_{a'_{u} \in \mathcal{A}_{u}}Q^{*}_{u}(s',a'_{u}).
\end{align}
%    The long term weighted reward can be defined as the weighted sum of the instantaneous reward over a $T \rightarrow \infty$ as follows
%    \begin{align}
%    \Phi=\sum_{t=0}^{T}\gamma^t \mathcal{R}
%    \end{align}
%
%    %$\mathcal{S}$ is the set of possible state. In each time slot the agent get following observes from the environment
%    %$s^{(i)}(t)=\{\gamma_{\text{UP}}^{(i)},\gamma_{\text{DL}}^{(i)}\}$
%
%    %$\mathcal{A}$ is the set of activities and decisions of uplink power allocation, DL beamforming, and scheduling.
%    %$A^{(i)}(t)=\{\boldsymbol{W}^{(i)},\boldsymbol{P}^{(i)},\boldsymbol{\rho}^{(i)}\}$
%
%
%    value-state function
%    \begin{align}
%    V(s,\pi)=E\Big[\sum\limits_{t=0}^{T-1}\gamma^t \mathcal{R}(s(t),\pi(t))|s(t)=s)\Big]
%    \end{align}
%
%    Bellman optimality equation
%    \begin{align}
%    V(s(t),\pi^*)=\max_{a\in \mathcal{A}}\Big[ E[\mathcal{R}(s(t), a(t))]+\gamma\sum P_{ss'}V(s'(t),\pi^*) \Big]
%    \end{align}
%
%
%    Optimal Q-value function $Q^*(s,a(t))$
%    \begin{align}
%    Q^*(s(t),a(t))=\Big[ E[\mathcal{R}(s(t), a(t))]+\gamma\sum P_{ss'}V(s'(t),\pi^*) \Big]
%    \end{align}
%
%    optimal policy can be obtained form
%    \begin{align}
%    V(s,\pi^*)=\max_{a(t)\in\mathcal{A}}Q^*(s(t),a(t))
%    \end{align}
%
However, in general, it is difficult to obtain the information about the transition probability $P_{ss’}(a_{u},\pi_{-u})$.~Nevertheless, the optimal strategy can be found through the available information recursively by using the Q-learning method.~Hence, the update of Q-value function is given by
\begin{align}\label{24}
Q_{u}(s,a_{u})=Q_{u}(s,a_{u})+\beta\Big[ \mathbb{E}[\mathcal{R}(s, a_{u},\pi_{-u})] +\gamma \max_{a'\in\mathcal{A}}Q_{u}(s',a'_{u})-Q_{u}(s,a_{u}) \Big],
\end{align}
where $\beta$ indicates the learning rate to determine the update of $Q_{u}(s,a_{u})$.

		\vspace{-5mm}
\subsection{Trajectory of UAVs-RISs Optimization as a Deep Q-Learning Problem}
\begin{figure}[t]
	\centering
	\vspace{-5mm}
	\includegraphics[width=3.5in]{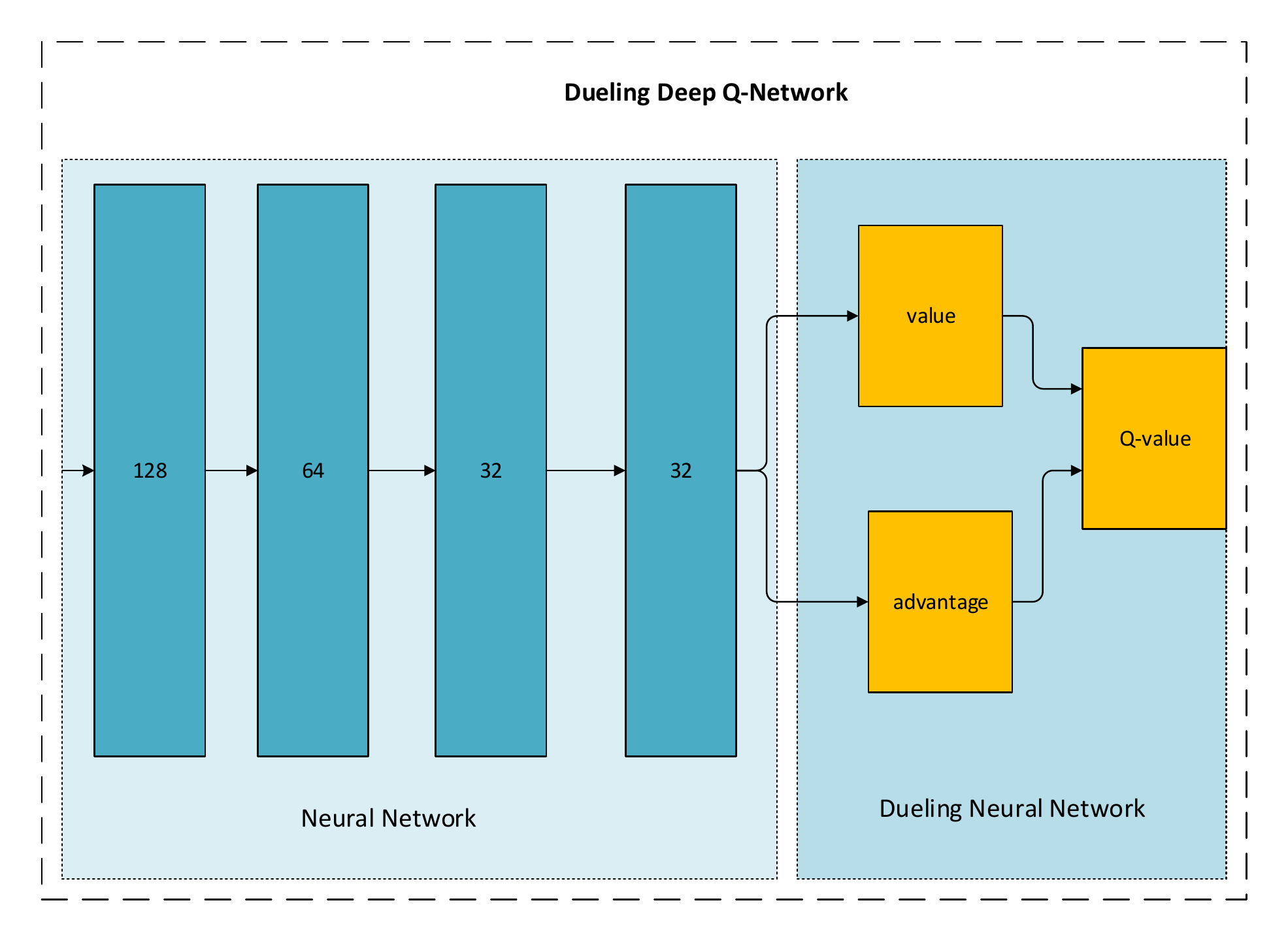}
		\vspace{-5mm}
	\caption{\small \textcolor{blue}{Dueling architecture used in our proposed solution.}}\label{fig44}
\end{figure}

In this section, the trajectory/velocity of each UAV, phase shifts of each RIS, and subcarrier allocation for MBS optimization problem is formulated as a Q-learning problem.~However, in huge Q value table, Q-learning can not be used.~Hence, we adopt multi-agent deep reinforcement learning where a deep neural network is employed to handle this issue\footnote{\textcolor{blue}{In our system model, each observation can be placed in a replay memory, and thus a mini-batch of replay memory would be selected. This mini-batch is used to train and update the weight of variables in the network. In fact, the weights of the neural network will be updated for each observation. Also, in each observation, each UAV modifies its location which leads to the change in channel gains.}}.~In this method,~Q-value function $Q_{u}(s,a_{u})$ is approximated through mapping an state to action.~The approximator function of the neural network is $Q_{u}(s,a_{u};\theta)\approxeq Q^{*}_{u}(s,a_{u})$,~where $\theta$ is an online network.~Accordingly, the updating Q-network for its weight to minimize the loss function can be represented by
\begin{align}
L_{u}(\theta)=\mathbb{E}_{s,a_{u},\nu{(s,a_{u}),s'}}\bigg[\bigg(y^{\text{DQN}}_{u}-Q_{u}(s,a_{u};\theta)\bigg)^{2}\bigg],
\end{align}
where $\nu(s,a_{u})=\mathbb{E}\Big[ \mathcal{R}\big(s,\pi_{u}(t),\pi^{*}_{-u}(t)\big)\Big]$.~Besides,  $y^{\text{DQN}}_{u}=\nu(s,a_{u})+\gamma \max_{a'_{u}\in \mathcal{A}_{u}Q_{u}}(s',a'_{u};\theta^{-})$, and $\theta^{-}$ illustrates the weights of a target network.
By using $\epsilon$-greedy policy, action $a_{u}$ can be obtained from the online network (ON) $Q_{u}(s,a_{u};\theta)$. The ON's weights are updated for some iterations, whereas the weights of the target network (TN) are fixed.
To overcome the unsteadiness of the learning procedure in DQN, we apply the experience replay strategy (ERS), where the experience replay memory $\mathcal{D}$ is employed to store the transition of $Q_{u}(s,a_{u};\theta)$. Rather than utilizing the current experience $Q_{u}(s,a_{u};\theta)$ during the learning procedure, sampling mini-batches of experiences can be adapted to train the TN from $\mathcal{D}$. The ERS guarantees that the optimal solution cannot be driven to a local minimum by decreasing the correlation among the training examples.
%~Besides, DDQN is employed to substitute the target $y^{\text{DQN}}_{u}$ with $y^{\text{DDQN}}_{u}$$=\nu(s,a_{u})+\gamma Q_{u}(s',\text{argmax}_{a'_{u}\in \mathcal{A}_{u}}Q_{u}(s',a'_{u};\theta^{-})$ since the Q-value function might be estimated over-optimistically when the same values are exploited to assess and choose an action in DQN and Q-learning approaches.
\textcolor{blue}{Furthermore, the dueling DQN is employed to obtain the estimation of the value function, $V(s)$, and the advantage function, $A(s,a_{u})=Q_{u}(s,a_{u})-V(s)$. In the dueling DQN, the last layer is split into two subnetworks to estimate $V(s)$ and $A(s,a_{u})$ separately. Therefore, the action value can be evaluated by combining $V(s)$ and $A(s,a_{u})$ which leads to a better policy evaluation \cite{Dueling}.~Please note that our Dueling DQN architecture includes an input layer with 32 neurons, four hidden layers (128, 64, 32, 32), and an output layer as can be seen in Fig. \ref{fig44}.} 
%Hence, the target $y^{DDQN}_{u}$ is given by
\begin{figure}[t]
	\centering
	\vspace{-5mm}
	\includegraphics[width=4.5in]{LDD.pdf}
	\vspace{-5mm}
	\caption{\small \textcolor{blue}{Flow chart of the proposed iterative algorithm.}}\label{fig3}
\end{figure}Fig. \ref{fig3}, shows the procedure of this strategy. In particular, the next state $s^{t+1}$ is used to estimate the optimal value of $Q_{u}(s^{t+1},a^{t+1}_{u};\theta)$ by ON and TN. Next, the target value $y^{\text{DQN}}_{u}$ can be achieved by adopting the current reward $\mathcal{R}$, and the discount factor $\gamma$. Finally, we can obtain the error value by subtracting the optimal value from the target value, which would be backpropagated in order to update the weights.  The multi-agent dueling DQN leaning algorithm is presented in Algorithm 1.
\textcolor{blue}{\begin{algorithm}[t]
		\caption{\textcolor{blue}{Multi-agent dueling DQN leaning method for the joint optimization of UAVs' trajectories, RISs' phase shifts, and subcarrier allocations}}
		\begin{algorithmic}[1]
			\renewcommand{\algorithmicrequire}{\textbf{Input:}}
			\renewcommand{\algorithmicensure}{\textbf{Output:}}
			\REQUIRE  List of allowed actions to be taken each UAV.\\
			\STATE \quad Initialize the replay memory $ D$, DQN network parameters $\theta$, and the target network replacement\\ \quad frequency $N^-$.
			\STATE \quad  Initialize the online network $Q(s; a;\theta )$ with weights $\theta$.
			\STATE \quad Initialize the target network $Q(s; a;\theta^- )$ with weights $\theta^- = \theta$.
			\STATE \quad Initialize the network state $s$.
			\STATE \quad\textbf{For} each steps  to $T$ steps \textbf{do}
			\STATE \quad\quad Each UAV chooses an action $a_i$ at state $s$ by using $\epsilon$-greedy policy from $Q_u(s; a_u;\theta )$.
			\STATE \quad\quad Each UAV gets the network state $s'$. Set $s'$ $\rightarrow$ $s$.
			\STATE \quad\quad Each UAV stores transition $(s; a_u; \mathcal{R}(s; a_u);s' )$ in $D$.
			\STATE \quad\quad Each UAV sets $y_u^{\text{DQN}}$ .
			\STATE \quad\quad Each UAV performs the gradient descent on $( y_u^{\text{DQN}} -Q_u(s; a_u;\theta ))^2$.
			\STATE \quad\quad In each $N^-$ step, each UAV replaces target parameters, $\theta^- = \theta$ .
			\STATE \quad    \textbf{End loop for} $T$ \textbf{steps}.
			\STATE   \textbf{Output} Optimal sequence of actions 
		\end{algorithmic}
\end{algorithm}}
\subsection{Active Beamforming and Subcarrier Allocation Design}
In this subsection, a suboptimal algorithm is introduced which has low complexity and can provide a locally optimal solution. Define $\mathbf{W}^{[l]}_{j,i(j)}(t)=\mathbf{w}^{[l]}_{j,i(j)}(t)\mathbf{w}^{[l],H}_{j,i(j)}(t)$ and $\mathbf{H}^{[l]}_{j,i(j)}(t)=\mathbf{h}^{[l]}_{j,i(j)}(t)\mathbf{h}^{[l],H}_{j,i(j)}(t)$. \textcolor{blue}{Then, the equivalent form of (6), (8), (9) on the $l$-th subcarrier can be rewritten as}
\begin{align}\label{sdp}
J^{[l]}_{i(j),i'(j)}({\mathbf{W}}(t))&=\log_2\bigg(1+\frac{\text{Tr}\big(\mathbf{H}^{[l]}_{j,i'(j)}(t){\mathbf{W}}^{[l]}_{j,i'(j)}(t)\big)}{I^{i'(j)}_{\text{CCI}}(t)+\text{Tr}\big(\mathbf{H}^{[l]}_{j,i'(j)}(t){\mathbf{W}}^{[l]}_{j,i(j)}(t)\big)+\sigma^{2}}\bigg)\\ \nonumber
&-\log_2\bigg(1+\frac{\text{Tr}(\mathbf{H}^{[l]}_{j,i(j)}(t){\mathbf{W}}^{[l]}_{j,i'(j)}(t))}{I^{i(j)}_{\text{CCI}}(t)+\text{Tr}(\mathbf{H}^{[l]}_{j,i(j)}(t){\mathbf{W}}^{[l]}_{j,i(j)}(t))+\sigma^{2}}\bigg)\leq 0,\\
R^{[l],\text{mmW}}_{i(j)}(t)&=z^{[l]}_{i(j),i'(j)}(t) \log_2\bigg(1+\frac{\text{Tr}(\mathbf{H}^{[l]}_{j,i(j)}(t){\mathbf{W}}^{[l]}_{j,i(j)}(t))}{I^{i(j)}_{\text{CCI}}(t)+\sigma^{2}}\bigg),\label{17}\\
R^{[l],\text{mmW}}_{i'(j)}(t)&=z^{[l]}_{i(j),i'(j)}(t)\log_2\bigg(1+\frac{\text{Tr}(\mathbf{H}^{[l]H}_{j,i'(j)}(t)\mathbf{W}^{[l]}_{j,i'(j)}(t))}{I^{i(j')}_{\text{CCI}}(t)+\text{Tr}(\mathbf{H}^{[l]H}_{j,i'(j)}(t)\mathbf{W}^{[l]}_{j,i(j)}(t))+\sigma^{2}}\bigg),
\end{align}
where $I^{i(j)}_{\text{CCI}}(t)=\sum_{j'\neq j}\sum_{k(j')}\text{Tr}\big(\mathbf{H}^{[l]}_{j',i(j)}(t){\mathbf{W}}^{[l]}_{j',k(j')}(t)\big).$
Furthermore, $C_1$ and $C_2$ can be, respectively, written as
\begin{align}
&C_{1}:~\sum_{l=1}^{L}\sum_{i=1}^{I}\sum_{i'=1}^{I} z^{[l]}_{i(j),i'(j)}(t)\left\{\text{Tr}(\mathbf{W}^{[l]}_{j,i(j)}(t))+\text{Tr}(\mathbf{W}^{[l]}_{j,i'(j)}(t))\right\}\leq P^{j}_{\max},\:\forall t, \:\text{if}:\:j\neq 0,\\
&C_{2}:~\sum_{i^{''}=1}^{I}\sum_{x=1}^{X} \rho^{[x]}_{i^{''}(j)}(t)\text{Tr}(\mathbf{W}^{[x]}_{0,i^{''}(j)}(t))\leq P^{0}_{\max},~\forall t, \:\text{if}:\:j= 0,
\end{align}where $\mathbf{W}^{[x]}_{0,i^{''}(j)}(t)=\mathbf{w}^{[x]}_{0,i^{''}(j)}(t)\mathbf{w}^{[x],H}_{0,i^{''}(j)}(t)$. To express the equivalent form of (\ref{rate2}), we first define $\mathbf{H}_u(t)=\sum_{u=1}^{U}\text{diag}(\mathbf{g}_{u,i^{''}(j)}^{[x],H}(t))\mathbf{G}^{[x]}_{0,u}(t)$ and $\mathbf{v}(t)=[e^{j\theta_{1}(t)},...,e^{j\theta_{N}(t)}]^H$. Therefore, the equivalent form of (\ref{rate2}) can be restated as
\begin{align}
R^{[x],\text{µW}}_{i^{''}(j)}(t)=\log_2\bigg(1+\frac{\text{Tr}(\rho^{[x]}_{i^{''}(j)}(t)\mathbf{W}^{[x]}_{0,i^{''}(j)}(t)\mathbf{H}^H_u(t)\mathbf{V}(t)\mathbf{H}_u(t))}{\delta^{2}}\bigg),
\end{align}where $\mathbf{V}(t)=\mathbf{v}(t)\mathbf{v}^H(t)$. Note that the auxiliary variable $\tilde{{\mathbf{W}}}^{[l]}_{j,i(j)}(t)=z^{[l]}_{i(j),i'(j)}(t)\mathbf{W}^{[l]}_{j,i(j)}(t)$ is introduced to overcome the product of a matrix variable and an integer variable in (\ref{17}). Accordingly, the big-M approach is exploited to decompose the multiplicative terms [53, page 361],\cite{Sun}. Thus, the product terms can be decomposed as follows:
\begin{align}
&C_{15}:~\tilde{\mathbf{W}}^{[l]}_{j,i(j)}(t)\preceq P^{j}_{\max}\mathbf{I}_\text{MC}z^{[l]}_{i(j),i'(j)}(t),\quad C_{16}:~\tilde{\mathbf{W}}^{[l]}_{j,i(j)}(t)\preceq{\mathbf{W}}_{j,i(j)}^{[l]}(t),\\
&C_{17}:~\tilde{\mathbf{W}}^{[l]}_{j,i(j)}(t)\succeq{\mathbf{W}}_{j,i(j)}^{[l]}(t)-(1-z^{[l]}_{i(j),i'(j)}(t))P^{j}_{\max}\mathbf{I}_\text{MC},\quad C_{18}:~\tilde{\mathbf{W}}^{[l]}_{j,i(j)}(t)\succeq\mathbf{0}.
\end{align}
Note that $C_{12}$ is the non-convex binary constraint which can be relaxed into a continuous interval between 0 and 1 expressed as
\begin{align}
&C_{12a}:~\sum_{j=1}^{J}\sum_{l=1}^{L}\sum_{i(j)=1}^{I}\sum_{i'(j)=1}^{I}z^{[l]}_{i(j),i'(j)}(t)-\sum_{j=1}^{J}\sum_{l=1}^{L}\sum_{i(j)=1}^{I}\sum_{i'(j)=1}^{I}\left( z^{[l]}_{i(j),i'(j)}(t)\right)^{2}\leq0,\text{}\label{b1}\\
&C_{12b}:~0 \leq z^{[l]}_{i(j),i'(j)}(t) \leq 1.\label{b2}
\end{align}
However, (\ref{b1}) is a reverse convex function. To handle it, we formulate problem (P1) as below:
\begin{align}\label{M_Uu}
\text{P2}:~&\underset{{\mathbf{z},\mathbf{w}}}{{\text{minimize}}}\:\:\frac{1}{T}\sum_{t=1}^{T} P_\text{total}(t)+\alpha\left(z^{[l]}_{i(j),i'(j)}(t)-\left( z^{[l]}_{i(j),i'(j)}(t)\right)^{2}\right)\\
&\text{s.t.}~
C_{1}-C_{4},\:C_{12b},\:C_{10}-C_{11},\: C_{15}-C_{18},\nonumber\\
&\quad~C_{19}:~\tilde{\mathbf{W}}^{[l]}_{j,i(j)}(t)\succeq\mathbf{0}, \nonumber\\
&\quad~C_{20}:~\text{Rank}(\tilde{\mathbf{W}}^{[l]}_{j,i(j)}(t))\leq 1, \nonumber
\end{align}
where $\alpha$ is a large penalty factor to penalize the objective function when $z^{[l]}_{i(j),i'(j)}$ does not take binary values. It can be shown that by choosing appropriate $\alpha$, (P1) and (P2) are equivalent and have the same results\footnote{\textcolor{blue}{For a sufficiently large value of $\alpha$, the integer variables are close to either $0$ or $1$. In particular, [60], [63]-[65] empirically showed that penalty factors with the value of $10^{(M+\log \frac{P_{\max}}{\sigma^{2}})}$, where $M$ is the number of antennas, work pretty well.}}.
Nevertheless, $C_{11}$ is non-convex since it is the difference of two logarithmic functions. To address it, we express $C_{11}$ as follows:
\begin{align}
&C_{11a}:~\log_2\bigg(1+\frac{\text{Tr}(\mathbf{H}^{[l]}_{j,i'(j)}(t)\tilde{\mathbf{W}}^{[l]}_{j,i'(j)}(t))}{I^{i'(j)}_{\text{CCI}}(t)+\text{Tr}(\mathbf{H}^{[l]}_{j,i'(j)}(t)\tilde{\mathbf{W}}^{[l]}_{j,i(j)}(t))+\sigma^{2}}\bigg)+\eta^{[l]}_{i(j),i'(j)}(t)\leq \log(1+\frac{P^{j}_{\max}}{\sigma^{2}}),\\
&C_{11b}:~\log_2\bigg(1+\frac{\text{Tr}(\mathbf{H}^{[l]}_{j,i(j)}(t)\tilde{\mathbf{W}}^{[l]}_{j,i'(j)}(t))}{I^{i(j)}_{\text{CCI}}(t)+\text{Tr}(\mathbf{H}^{[l]}_{j,i(j)}(t)\tilde{\mathbf{W}}^{[l]}_{j,i(j)}(t))+\sigma^{2}}\bigg)+\eta^{[l]}_{i(j),i'(j)}(t)\geq \log(1+\frac{P^{j}_{\max}}{\sigma^{2}}),
\end{align}
where $\eta^{[l]}_{i(j),i'(j)}(t)$ is the slack variable. To further simplify, (P2) can be rewritten as
\begin{align}\label{p3}
\text{P3}:~&\underset{{\mathbf{z},\mathbf{w},\boldsymbol{\eta}(t)}}{{\text{minimize}}}\:\frac{1}{T}\sum_{t=1}^{T} P_\text{total}(t)+\alpha\left(z^{[l]}_{i(j),i'(j)}(t)-\left( z^{[l]}_{i(j),i'(j)}(t)\right)^{2}\right)\\
&\text{s.t.}~C_{1}-C_{4},\:C_{12b},\:C_{10},\: C_{11a}-C_{11b},\:C_{15}-C_{20},\nonumber\\
&\quad~C_{15}:~\sum_{l=1}^{L}\sum_{x=1}^{X}F(\tilde{\mathbf{W}}(t))-G(\tilde{\mathbf{W}}(t))\geq R^{i}_{\min}, \quad~C_{16}:~\sum_{l=1}^{L}T(\tilde{\mathbf{W}}(t))-R(\tilde{\mathbf{W}}(t))\geq \tilde{R}^{i}_{\min},\nonumber\\
&\quad~C_{17}:~A^{[l]}_{i(j),i'(j)}(\tilde{\mathbf{W}}(t),\boldsymbol{\eta}(t))-B^{[l]}_{i(j),i'(j)}(\tilde{\mathbf{W}}(t))\leq 0,\nonumber\\
&\quad~C_{18}:~D^{[l]}_{i(j),i'(j)}(\tilde{\mathbf{W}}(t))-C^{[l]}_{i(j),i'(j)}(\tilde{\mathbf{W}}(t),\boldsymbol{\eta}(t))\leq 0,\nonumber
\end{align}
where $\boldsymbol{\eta}(t)$ is the collection of optimization variables $\eta^{[l]}_{i(j),i'(j)}(t)$ and
\begin{align}
F(\tilde{\mathbf{W}}(t))&=\log_2\bigg(I^{i(j)}_{\text{CCI}}(t)+\sigma^{2}+{\text{Tr}(\mathbf{H}^{[l]}_{j,i(j)}(t)\tilde{\mathbf{W}}^{[l]}_{j,i(j)}(t))}\bigg)\nonumber\\&+\log_2\bigg(\text{Tr}(\mathbf{H}^{[l]H}_{j,i'(j)}(t)\mathbf{W}^{[l]}_{j,i'(j)}(t))+I^{i(j')}_{\text{CCI}}(t)+\text{Tr}(\mathbf{H}^{[l]H}_{j,i'(j)}(t)\mathbf{W}^{[l]}_{j,i(j)}(t))+\sigma^{2}\bigg)\nonumber\\&+\log_2\bigg(\text{Tr}(\rho^{[x]}_{i^{''}(j)}(t)\mathbf{W}^{[x]}_{0,i^{''}(j)}(t)\mathbf{H}^H_u(t)\mathbf{V}(t)\mathbf{H}_u(t))+\delta^{2}\bigg),
\end{align}
\begin{align}
G(\tilde{\mathbf{W}}(t))&=\log_2\left(I^{i(j')}_{\text{CCI}}(t)+\text{Tr}(\mathbf{H}^{[l]H}_{j,i'(j)}(t)\mathbf{W}^{[l]}_{j,i(j)}(t))+\sigma^{2}\right)+\log_2(\delta^{2})\nonumber\\&+\log_2\left(I^{i(j)}_{\text{CCI}}(t)+\sigma^{2}\right),\\
T(\tilde{\mathbf{W}}(t))&=\log_2\bigg(\text{Tr}(\mathbf{H}^{[l]H}_{j,i'(j)}(t)\mathbf{W}^{[l]}_{j,i'(j)}(t))+I^{i(j')}_{\text{CCI}}(t)+\text{Tr}(\mathbf{H}^{[l]H}_{j,i'(j)}(t)\mathbf{W}^{[l]}_{j,i(j)}(t))+\sigma^{2}\bigg),\\
R(\tilde{\mathbf{W}}(t))&=\log_2\left(I^{i(j')}_{\text{CCI}}(t)+\text{Tr}(\mathbf{H}^{[l]H}_{j,i'(j)}(t)\mathbf{W}^{[l]}_{j,i(j)}(t))+\sigma^{2}\right),\\
A^{[l]}_{i(j),i'(j)}(\tilde{\mathbf{W}}(t),\boldsymbol{\eta}(t))&=\log_2\bigg({I^{i'(j)}_{\text{CCI}}(t)+\text{Tr}(\mathbf{H}^{[l]}_{j,i'(j)}(t)\tilde{\mathbf{W}}^{[l]}_{j,i(j)}(t))+\sigma^{2}}+{\text{Tr}(\mathbf{H}^{[l]}_{j,i'(j)}(t)\tilde{\mathbf{W}}^{[l]}_{j,i'(j)}(t))}\bigg)\nonumber\\&+\eta^{[l]}_{i(j),i'(j)}(t)- \log(1+\frac{P^{j}_{\max}}{\sigma^{2}}),	\\
B^{[l]}_{i(j),i'(j)}(\tilde{\mathbf{W}}(t))&=\log_2\bigg({I^{i(j)}_{\text{CCI}}(t)+\text{Tr}(\mathbf{H}^{[l]}_{j,i'(j)}(t)\tilde{\mathbf{W}}^{[l]}_{j,i(j)}(t))+\sigma^{2}}\bigg),\\
C^{[l]}_{i(j),i'(j)}(\tilde{\mathbf{W}}(t),\boldsymbol{\eta}(t))&=\log_2\bigg({I^{i(j)}_{\text{CCI}}(t)+\text{Tr}(\mathbf{H}^{[l]}_{j,i(j)}(t)\tilde{\mathbf{W}}^{[l]}_{j,i(j)}(t))+\sigma^{2}}+{\text{Tr}(\mathbf{H}^{[l]}_{j,i(j)}(t)\tilde{\mathbf{W}}^{[l]}_{j,i'(j)}(t))}\bigg)\nonumber\\&+\eta^{[l]}_{i(j),i'(j)}(t)- \log(1+\frac{P^{j}_{\max}}{\sigma^{2}}),\\
D^{[l]}_{i(j),i'(j)}(\tilde{\mathbf{W}}(t))&=\log_2\bigg({I^{i(j)}_{\text{CCI}}(t)+\text{Tr}(\mathbf{H}^{[l]}_{j,i(j)}(t)\tilde{\mathbf{W}}^{[l]}_{j,i(j)}(t))+\sigma^{2}}\bigg).
\end{align}
As can be observed, (P3) can be considered as a difference of convex function program. Consequently, a locally optimal solution can be obtained by employing successive convex approximation (SCA) \cite{Diehl}.~\textcolor{blue}{To obtain an initial point, we resort to the previous works \cite{TSP_Ata,I1,I2}}. Since $G(\tilde{\mathbf{W}}(t))$, $R(\tilde{\mathbf{W}}(t))$,  $B^{[l]}_{i(j),i'(j)}(\tilde{\mathbf{W}}(t))$, and $D^{[l]}_{i(j),i'(j)}(\tilde{\mathbf{W}}(t))$ are differentiable convex functions, we can apply first order-Taylor approximation as follows:\footnote{For notational simplicity, we remove the subscriber from the beamforming vector.}
\begin{align}
G(\tilde{\mathbf{W}}(t))&\leq G(\tilde{\mathbf{W}}^{(k)}(t))+\text{Tr}\bigg(\nabla _{\tilde{{{\mathbf{W}}}}(t)}G(\tilde{\mathbf{W}}^{(k)}(t))^T\big(\tilde{\mathbf{W}}(t)-\tilde{\mathbf{W}}^{(k)}(t)\big)\bigg)\triangleq\bar{G}(\tilde{\mathbf{W}}(t)),\\
R(\tilde{\mathbf{W}}(t))&\leq \hat{R}(\tilde{\mathbf{W}}^{(k)}(t))+\text{Tr}\bigg(\nabla _{\tilde{{{\mathbf{W}}}}(t)}\hat{R}(\tilde{\mathbf{W}}^{(k)}(t))^T\big(\tilde{\mathbf{W}}(t)-\tilde{\mathbf{W}}^{(k)}(t)\big)\bigg)\triangleq\bar{R}(\tilde{\mathbf{W}}(t)),\\
B^{[l]}_{i(j),i'(j)}(\tilde{\mathbf{W}}(t))&\leq B^{[l]}_{i(j),i'(j)}(\tilde{\mathbf{W}}^{(k)}(t))+\text{Tr}\bigg(\nabla _{\tilde{{{\mathbf{W}}}}(t)}B^{[l]}_{i(j),i'(j)}(\tilde{\mathbf{W}}^{(k)}(t))^T\big(\tilde{\mathbf{W}}(t)-\tilde{\mathbf{W}}^{(k)}(t)\big)\bigg)\nonumber\\&\triangleq\bar{B}^{[l]}_{i(j),i'(j)}(\tilde{\mathbf{W}}(t)),\\
D^{[l]}_{i(j),i'(j)}(\tilde{\mathbf{W}}^{(k)}(t))&\leq D^{[l]}_{i(j),i'(j)}(\tilde{\mathbf{W}}^{(k)}(t))+\text{Tr}\bigg(\nabla _{\tilde{{{\mathbf{W}}}}(t)}D^{[l]}_{i(j),i'(j)}(\tilde{\mathbf{W}}^{(k)}(t))^T\big(\tilde{\mathbf{W}}(t)-\tilde{\mathbf{W}}^{(k)}(t)\big)\bigg)\nonumber\\&\triangleq\bar{D}^{[l]}_{i(j),i'(j)}(\tilde{\mathbf{W}}^{(k)}(t)).
\end{align}
For any given feasible point $\tilde{\mathbf{W}}^{(k)}(t)$, we have the relaxed problem
\begin{align}\label{p4}
\text{P4}:~&\underset{\mathbf{z},\mathbf{w},\boldsymbol{\eta}(t)}{{\text{minimize}}}\:\frac{1}{T}\sum_{t=1}^{T} P_\text{total}(t)+\alpha\left(z^{[l]}_{i(j),i'(j)}(t)-\left( z^{[l]}_{i(j),i'(j)}(t)\right)^{2}\right)\nonumber\\
&\text{s.t.}~
C_{1}-C_{4},\:C_{12b},\:C_{10},\:C_{11a}-C_{11b},\: C_{15}-C_{19},\nonumber\\
&\quad~C_{15}:~\sum_{l=1}^{L}\sum_{x=1}^{X}F(\tilde{\mathbf{W}}(t))-\bar{G}(\tilde{\mathbf{W}}(t))\geq R^{i}_{\min},\quad~C_{16}:~\sum_{l=1}^{L}T(\tilde{\mathbf{W}}(t))-\bar{R}(\tilde{\mathbf{W}}(t))\geq \tilde{R}^{i}_{\min},\nonumber\\
&\quad~C_{17}:~A^{[l]}_{i(j),i'(j)}(\tilde{\mathbf{W}}(t),\boldsymbol{\eta}(t))-\bar{B}^{[l]}_{i(j),i'(j)}(\tilde{\mathbf{W}}(t))\leq 0,\nonumber\\
&\quad~C_{18}:~\bar{D}^{[l]}_{i(j),i'(j)}(\tilde{\mathbf{W}}(t))-C^{[l]}_{i(j),i'(j)}(\tilde{\mathbf{W}}(t),\boldsymbol{\eta}(t))\leq 0.
\end{align}
Problem (P4) is a standard SDP that can be solved by dropping the rank one constraint $C_{20}$.~\textcolor{blue}{By dropping the rank-one constraint, there is no guarantee to obtain an optimal solution. In such a case, the Gaussian randomization method can be exploited to obtain a near-to-optimal solution.} 
\vspace{-5mm}
\section{Computational Complexity}
\vspace{-2mm}
In this section, we evaluate the computational complexity of the proposed scheme.~For the first subproblem, in DNN, let denote $K$, $\phi_{0}$, and $\phi_{k}$ as the training layers, the size of the input layer that is related to the number of states, and the number of neurons in the $k$ layer, respectively.~As a result, the computational complexity for each agent in each time step is $\mathcal{O}(\phi_{0}\phi_{k}+\sum_{k=1}^{K-1} \phi_{k}\phi_{k+1})$.
Furthermore, in the training phase, each mini-batch has $E^{\text{episode}}$ episodes with each episode being $T$ time steps, each trained model is completed over $I$ iterations until convergence.
Consequently, the total computational complexity in DNN is $\mathcal{O}(IE^{\text{episode}}T(\phi_{0}\phi_{k}+\sum_{k=1}^{K-1} \phi_{k}\phi_{k+1}))$. Also, for DQN the complexity order is $\mathcal{O}(|\mathcal{S}|^{2}\times \mathcal{A})$. Accordingly, the complexity order of DQN algorithm is $\mathcal{O}\bigg(IE^{\text{episode}}T(\phi_{0}\phi_{k}+\sum_{k=1}^{K-1} \phi_{k}\phi_{k+1})+(|\mathcal{S}|^{2}\times \mathcal{A})\bigg)$.~For the second subproblem, it can be observed that (P4) includes $LJI$ variables and $J+IJ+LJ+2LIJ+4LIJ+LIJ$ linear convex constraints.~It can be concluded that the complexity order of SCA is $\mathcal{O}(LJI)^{2}(J+IJ+LJ+7LIJ)$.~Consequently, the complexity order for the second subproblem can be estimated asymptotically as $\mathcal{O}(LJI)^{3}$ which indicates polynomial time complexity.

  \begin{table}
	\renewcommand{\arraystretch}{1.05}
	\centering
	\caption{\small Simulation Parameters}
	\label{table-notations}
	\begin{tabular}{| c| c| c| c|}
		%\begin{tabular}{| c| l| }
		\hline
		\textbf{Parameters}& \textbf{Values}&\textbf{Parameters}& \textbf{Values}\\\hline
		%%%%%%%%%%%parameters%%%%%%%%%%%%%%%%%%
		Cell radius &  $500$ m &Number of UAVs-RISs,~$U$ &$2$\\  \hline
		Path loss exponent  & $2$&Number of mmW subcarriers,~$L$&16 \\ \hline
		Rician factor,~$K_r$ & $5$~dB &Number of µW subcarriers,~$X$&8\\ \hline
		Antenna noise power for small-cells,~$\xi^{[l],\text{mmW}}$ & $-120$ dBm &Penalty factor,~$\alpha$ & $10^{5}$ \\ \hline
		 Antenna noise power for macro-cell,~$\xi^{[x],\text{µW}}$ & $-120$ dBm&Maximum transmit power at each SBS, $P^j_\text{max}$ & $42$ dBm \\ \hline
		Carrier frequency,~$f_c$ & $1.5$ GHz&Maximum transmit power at each MBS, $P^0_\text{max}$ &$30$ dBm\\ \hline
		Number of antenna at the MBS,~$M_{\text{mc}}$ & $8$&Minimum QoS of user $i$ in macro-cell, $R^i_\text{min}$ & $1$ bps/Hz \\ \hline
		Number of antenna at the SBS,~$M_{\text{sc}}$ & $4$&Minimum QoS of user $i$ in small-cell, $\tilde{R}^i_\text{min}$& $1$ bps/Hz \\ \hline
		Number of reflecting elements,~$N$ & $8$&Minimum and maximum flight altitude, $h_\text{min}$ and  $h_\text{max}$&$30-100$ m\\ \hline
		Number of users in macro-cell,~$I$ & $4$&Maximum flight velocity, $V_\text{u,max}$& $10$ m/s\\ 	 \hline
		Number of users in $j$-th small-cell,~$I(J)$ & $3$&Maximum acceleration of each UAV, $a_u$&$1$~m$/\text{s}^{2}$ \\ \hline
	\end{tabular}
\end{table}

\section{Simulation Results}\label{simulation}
In this section, numerical results are presented to assess the performance of the considered HetNet consisting of one MBS coexisting with 3 SBSs and 2 UAVs-RISs. A three-dimensional (3D) environment is considered where a uniform linear array (ULA)
and a uniform rectangular array (URA) are used on the MBS and each RIS, respectively. It is assumed that all users are uniformly distributed in a geographical area. The UAVs-RISs and SBSs are randomly deployed on the macro-cell to serve multiple users. \textcolor{blue}{The simulation parameters are presented in Table I unless otherwise is specified.} To investigate the effectiveness of the proposed algorithm 
%in the PD-NOMA-based NetNet consisted of multiple UAVs-RISs, 
four schemes are introduced as follows: MC-PD-NOMA-based HetNet with multiple UAVs-RISs (Proposed algorithm); MC-PD-NOMA-based HetNet without UAVs-RISs (Baseline scheme 1); OMA-based HetNet with multiple UAVs-RISs (Baseline scheme 2); OMA-based HetNet without UAVs-RISs (Baseline scheme 3).
\begin{figure}[!tbp]
	\centering
	\subfloat[Transmit power of the BSs versus \protect\\minimum data rate.]{\includegraphics[width=0.5\textwidth]{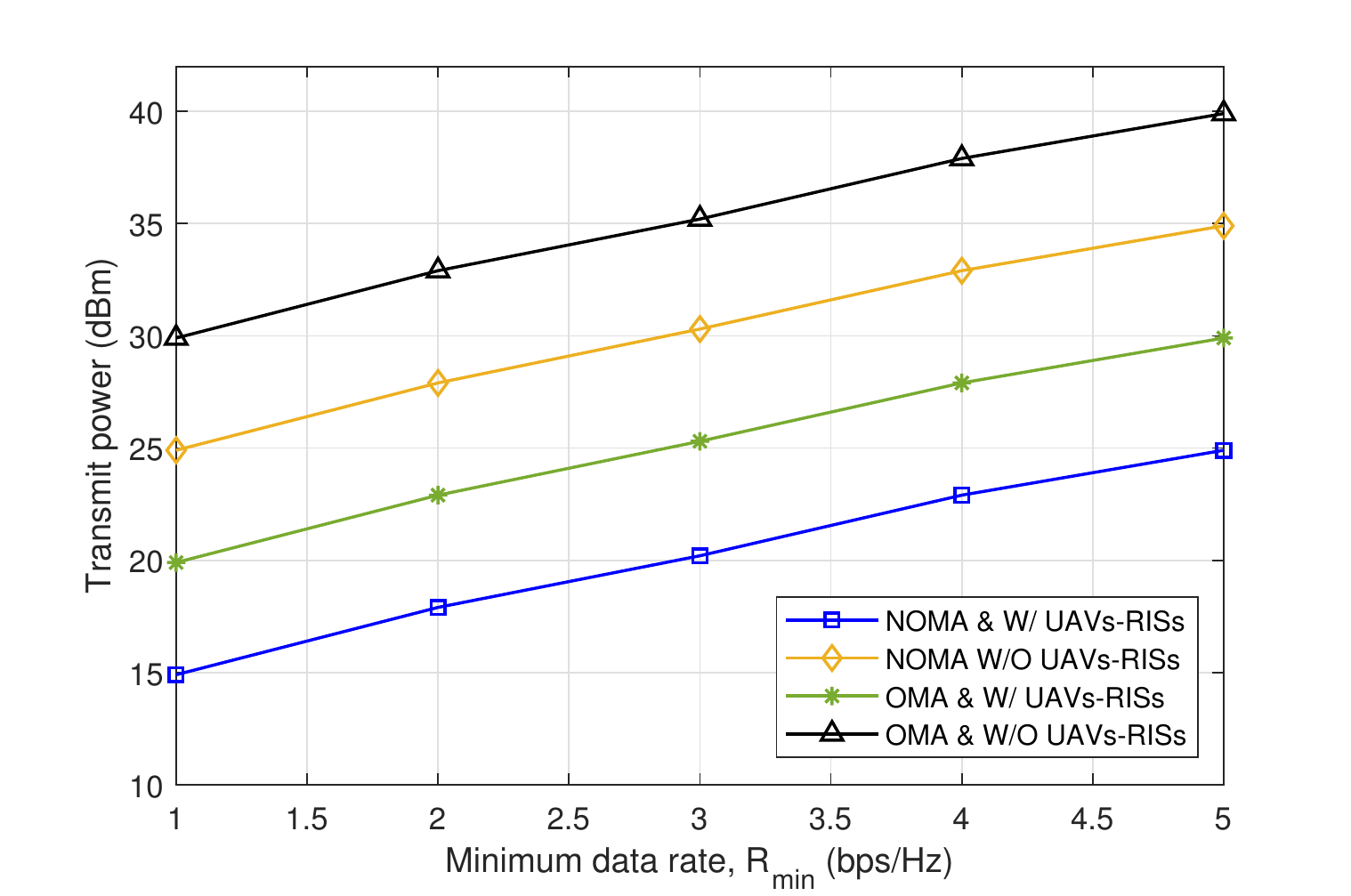}\label{fig:f1}}
	\hfill
	\centering	\subfloat[Transmit power of the BSs versus \protect\\minimum data rate]{\includegraphics[width=0.45\textwidth]{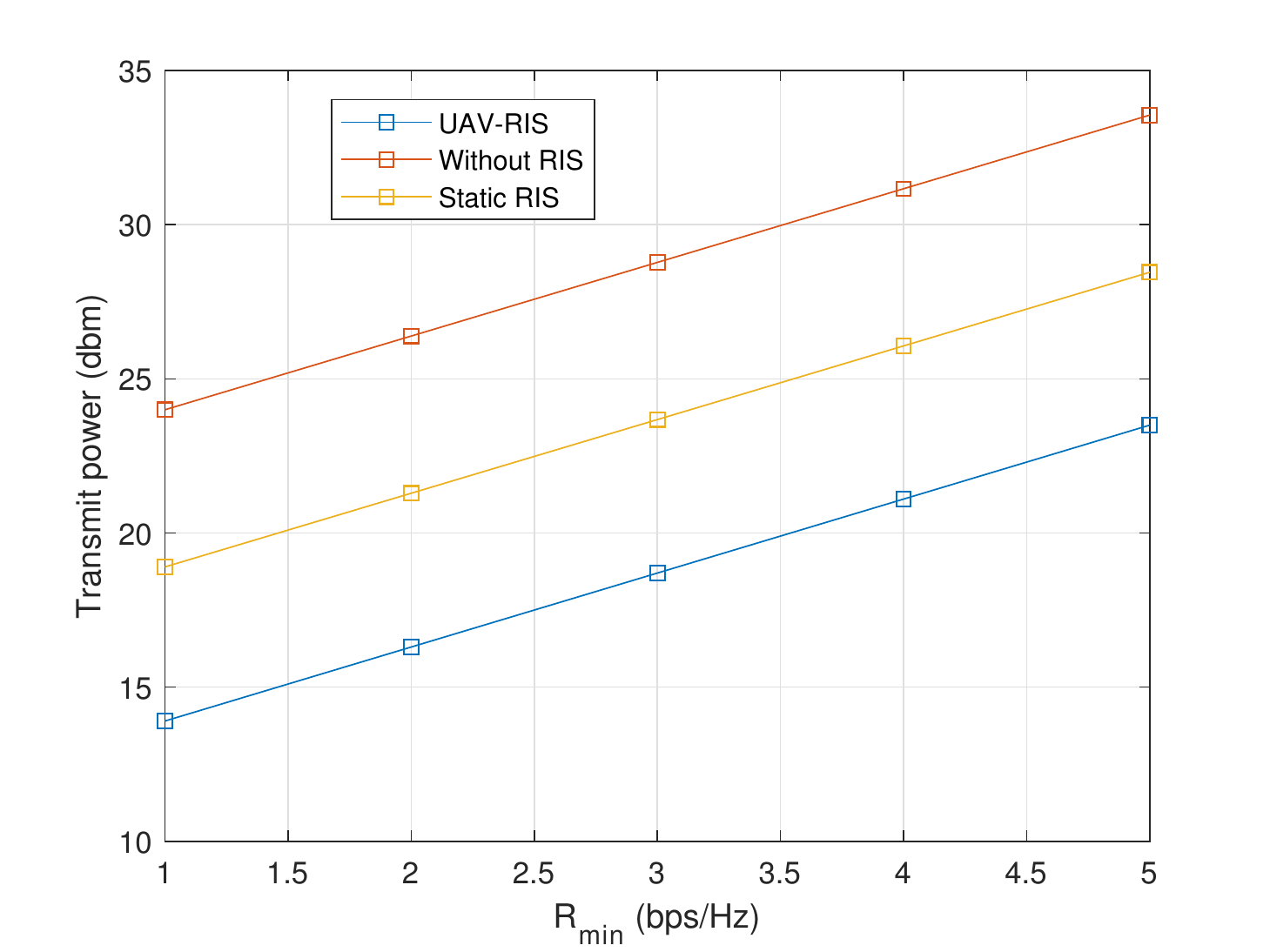}\label{fig:ff1}}
	\caption{\textcolor{blue}{\small Transmit power of the BSs versus minimum data rate.}}
	\centering
	{\includegraphics[width=0.5\textwidth]{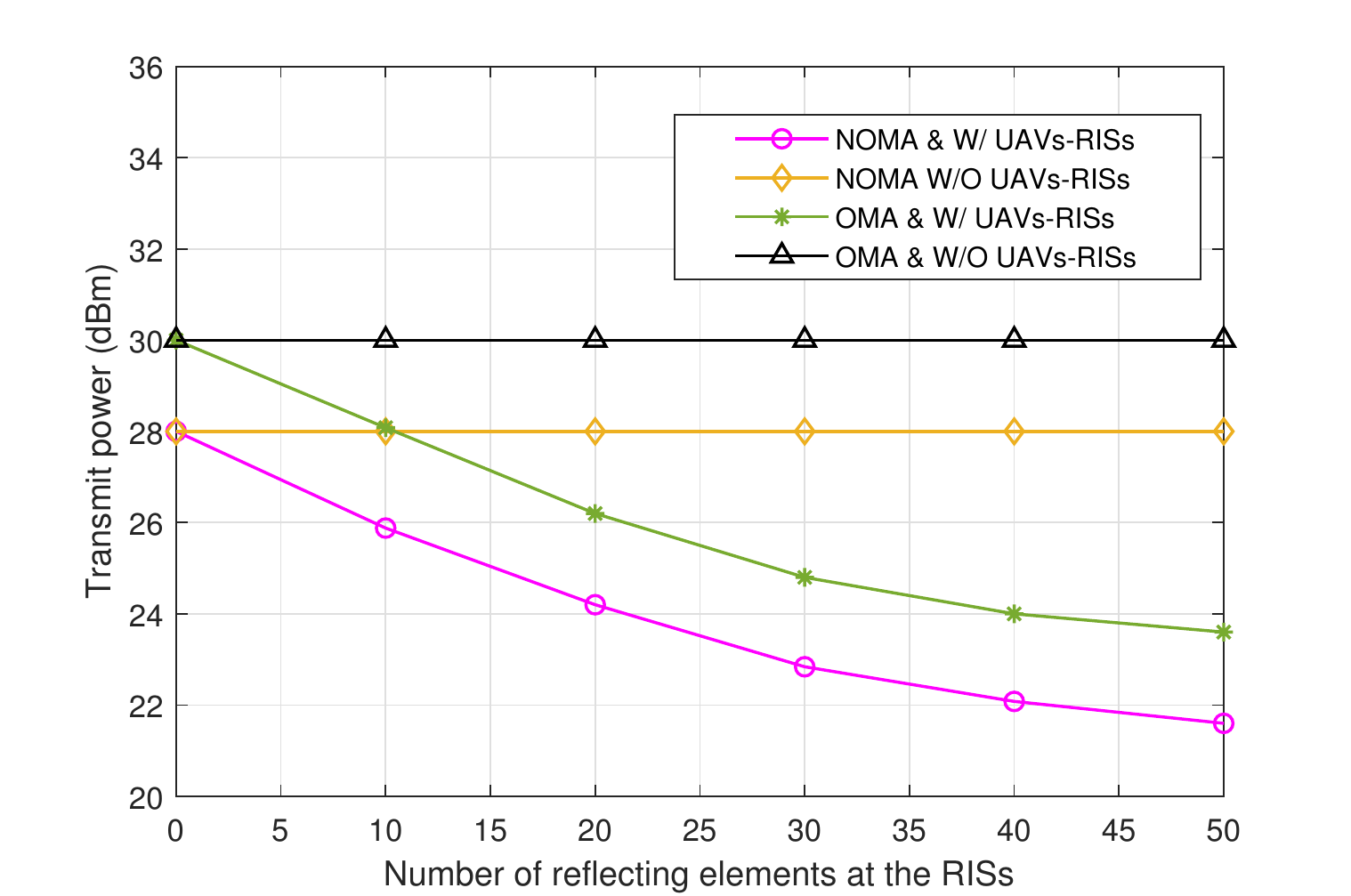}\label{fig:f2}}
	\caption{Transmit power of the BSs versus number of  reflecting  elements at the RIS.}
	\hfill
\end{figure}
%\begin{figure}[t]
%	\centering
%	\includegraphics[width=4.5in] {Fig_R_3.eps}
%	\caption{\small Total power consumption of the BSs versus minimum data rate.}\label{figure_N}
%\end{figure}
%Fig. 4 shows the total power consumption versus the number of users for different schemes. It can be observed that for the proposed scheme and baseline schemes the total power consumption increases as the number of users increases as well.~Since for a higher number of users, more degrees of freedom (DoF) at the BSs are used to ensure the QoS requirements of all users. In fact, the BSs are required to steer the beams towards the users to meet their QoS requirement which increases the transmit power and the total power consumption subsequently. In contrast, the proposed scheme can fulfill the stringent QoS constraints and obtain a much less transmit power compared to the baseline schemes, because of the proposed optimization.~Furthermore, each UAV-RIS has to devote more DoF which reduces the flexibility in trajectory and phase passive beamforming at each UAV-RIS, thus it degrades the performance concerning the total power consumption.~This figure also demonstrates the effect of UAVs-RISs on reducing the total power transmission which helps each BS to consume much small power transmission for supporting the users.~It is concluded that the UAVs-RISs can help the HetNet to diminish the transmit power of each BS significantly even for the SBSs operating on the NOMA scheme.
\vspace{-5mm}
\subsection{Transmit Power Versus Minimum Data Rate}
In Fig. \ref{fig:f1}, we plot the transmit power versus the minimum data rate requirements for different baseline schemes.~As expected, the total power begins to increase gradually as the minimum data rate requirement increases.~In particular, as the data rate requirements of the users become more stringent, a higher transmit power is required to fulfill the requirements, which increases the total power of the system significantly.~In other words, when the minimum required data rate increases, users with low channel qualities need more transmit power of each BS to meet their QoS requirement.~\textcolor{blue}{This also affects the UAVs-RISs which reduce the flexibility of the trajectory and yield to transmit power increase at the MBS.} As can be observed, the proposed algorithm with NOMA and UAVs-RISs provides notable power savings compared to other baseline schemes. However, for high data rate requirements, transmit power is more sensitive to variations in the data rate requirements, which leads to a faster increase in total power. It can also be observed that by increasing the data rate requirements of users, the NOMA scheme requires less transmit power as compared to the OMA scheme with the presence of multiple UAVs-RISs, which depicts the superiority of NOMA over the OMA scheme. Besides, deploying RISs in HetNet can reduce the power by helping the MBS in reflecting the incident signals. Thus, the RIS is a promising technology that can bring significant gain in terms of transmit power compare to the case where no RIS is used. On the other hand, since each RIS has a fixed position (e.g., on the walls, buildings, and so on), it can limit the performance of the system, especially when there is a non-LoS link between the RIS and each user. Accordingly, UAVs can play an important role in producing the LoS links by carrying RISs in the sky. Hence, the UAVs-RISs can substantially assist the considered system in reducing transmission power.~\textcolor{blue}{This figure also compares the UAV-RISs to the other scenarios namely Without UAV-RIS, in which no UAV-RIS is deployed, static RIS, in which the locations of RISs are fixed without any mobility.~This figure shows how much improvement can be obtained via UAV-RISs compared to a conventional one that does not exploit any UAV-RIS.~Also, this figure compares with static RIS and investigates how much improvement UAV brings to the system as compared to a static RIS.}

		\vspace{-3mm}
\subsection{Transmit Power Versus Number of Reflecting Elements at the RISs}
Fig. \ref{fig:f2} demonstrates the advantage of phase shifts at the RISs, which incurs transmission power reduction as compared to the case where no RIS is employed. This figure shows that the transmit power for all schemes reduces as the number of reflecting elements at the RISs increases. In the non-RIS case, as expected, using the NOMA scheme as a multiple access technique has a high-performance gain as compared to OMA schemes since more degrees of freedom (DoF) are available for resource allocator to support more users. However, the proposed scheme with RISs provides considerable performance gain as compared to the non-RIS case, which reveals the benefit of deploying RIS with low-cost phase shifts. As can be observed, the UAVs can make the resource allocation more flexible by optimizing their trajectories to support users with severe channel consideration. On the other hand, by deploying the UAVs-RISs, passive elements at the RISs and UAVs' trajectories can jointly be optimized to reduce the transmit power significantly.
\begin{figure}[!tbp]
	\centering
	\subfloat[Transmit power of the BSs versus the number of antennas\protect\\ at the MBS.]{\includegraphics[width=0.5\textwidth]{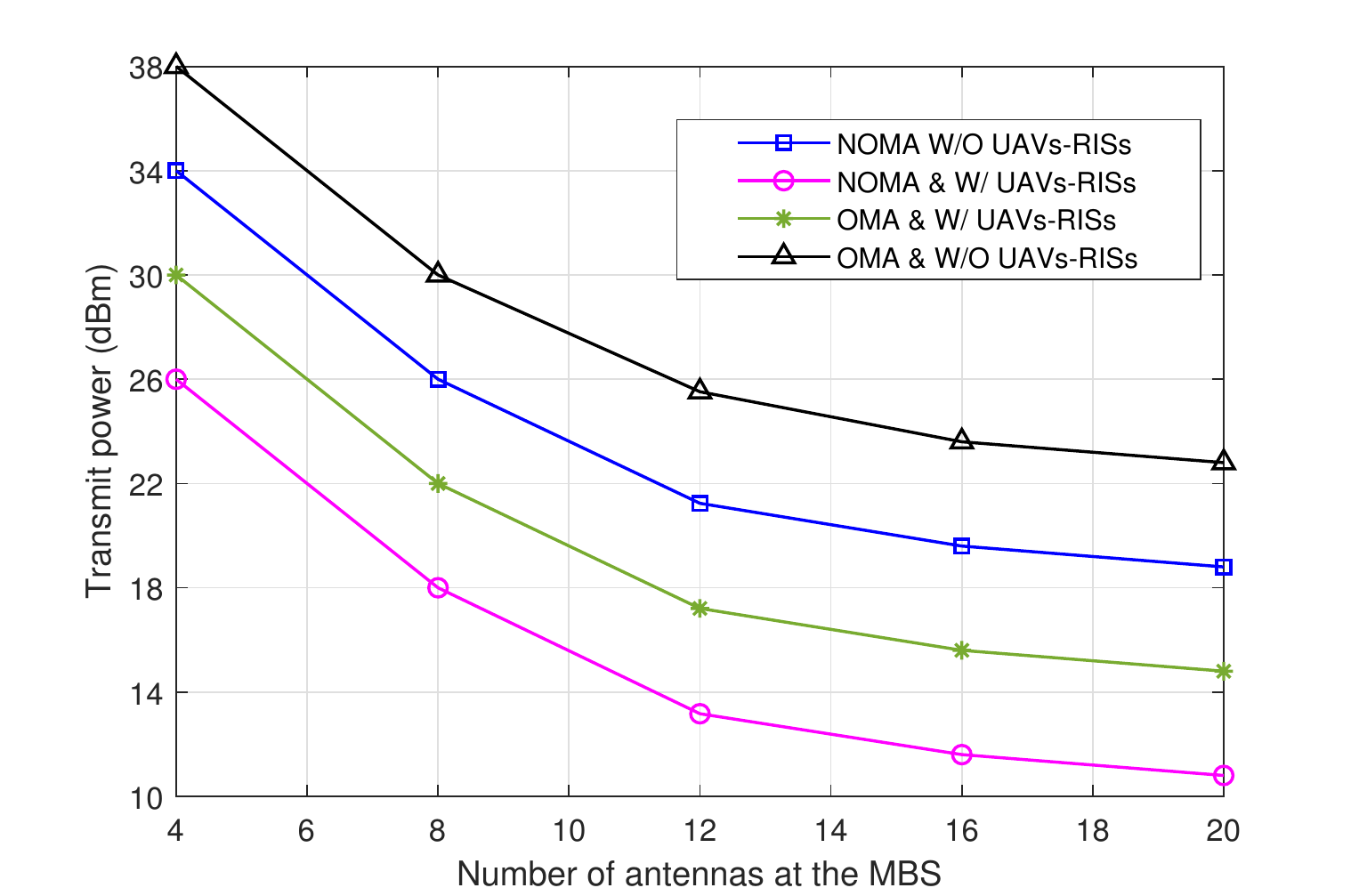}\label{fig:f3}}
	\hfill
	\subfloat[Convergence behavior of the proposed resource allocation algorithm.]{\includegraphics[width=0.45\textwidth]{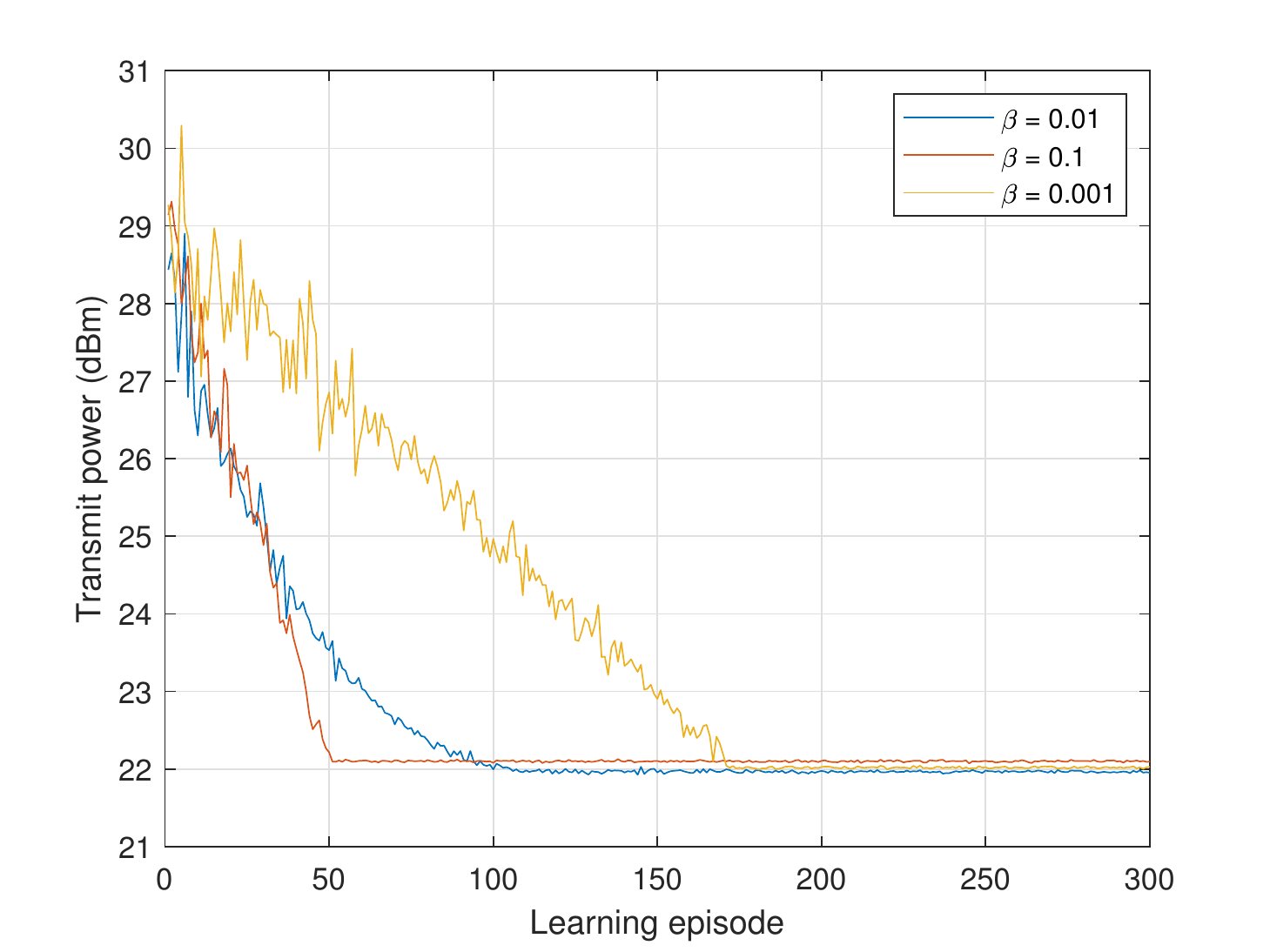}\label{fig:f4}}
	\caption{\small Transmit power of the BSs versus the number of antennas at the MBS and convergence behavior.}
\end{figure}		\vspace{-10mm}
\subsection{Transmit Power Versus Number of Antennas at the MBS}
Fig. \ref{fig:f3} shows the effect of number of transmit antennas at the MBS for different baseline schemes. As expected, by increasing the number of the transmit antennas at the MBS, less transmit power is required for information transmission. This is because by increasing transmit antennas, more DoFs are available for spatial multiplexing, which accordingly reduces the transmission power at the MBS. It can also be observed that by adopting the phase shifts at the RISs and UAVs' trajectories, more performance gain can be obtained compared to the case with NOMA/OMA without the UAVs-RISs. On the other hand, the NOMA scheme achieves a better performance than the OMA scheme, indicating the advantage of NOMA in reducing the transmit power of the system. It is concluded that the multiple antenna technique can efficiently decrease the transmission power of the system in different transmission policies.

%\begin{figure}[t]
%	\centering
%	\includegraphics[width=4.5in] {Fig_R_2.eps}
%	\caption{\small Total power consumption of the BSs versus the number of antennas at the MBS.}\label{figure_R}
%\end{figure}
%\begin{figure}[t]
%	\centering
%	\includegraphics[width=4.5in] {NReward_Leraning_Episode.eps}
%	\caption{\small Convergence behavior of the proposed resource allocation algorithm.}\label{Convergence}
%\end{figure}
		\vspace{-3mm}
\subsection{Convergence Behavior}
Fig. \ref{fig:f4} plots the transmit power per episode for our proposed resource allocation policy.~This figure illustrates that the transmit power gradually decreases as training continues, demonstrating the effectiveness of the proposed training approach. When the training episode approximately reaches 300, the performance gradually converges despite some fluctuations due to mobility-induced channel fading in mobile environments and the adopted greedy policy for the exploration and the exploitation in the training process. It can be seen that the dimension of action and state space for our proposed algorithm are large, so the proposed learning approach needs almost 300 training episodes to converge appropriately.~This figure also demonstrates the convergence of our proposed algorithm for different learning rate $\beta$.~We can observe that when $\beta$ increases from $0.001$ to $0.01$, the reward value improves by $\%1.05$ after convergence.~However, when $\beta$ increases to $0.1$, the reward converges very quick.~We can conclude that the learning rates are sensitive to the convergence rate and need to be appropriately selected.	
%	\vspace{-8mm}
\subsection{Trajectory}
Fig. 7 represents the UAVs' trajectories for the proposed
algorithm with three SBSs and two UAVs-RISs. It can be observed that each UAV-RIS flies straightly from the initial point toward
the final point with optimized altitude and flight velocity under two scenarios with different initial positions. Specifically, due to the minimum QoS requirement, each UAV-RIS is required to control its flight velocity and trajectory in order to establish a strong channel gain with each ground user. However, it can cost a high flight power consumption. This power consumption can be diminished by mounting the RIS under each UAV. Another interesting observation is that in the first point, two UAVs-RISs serves one small-cell and then quickly fly toward the other two small-cells. When each UAV-RIS comes close to any desired user, it slows down its velocity to promote effective information communications. Thus, the RIS along with the UAV can yield higher flexibility in designing UAV's trajectory which can decrease the flight power consumption and help serving cellular users with NLoS links.~Furthermore, it can be perceived that the final positions of the UAVs in both scenarios do not depend on the initial positions of the UAVs which investigates the efficiency of the proposed algorithm.~\textcolor{blue}{It should be noted that by considering fixed mobility for each user, trajectory optimization turns into location optimization. On the other hand, considering the movements of each UAV-RIS location during the time slots leads to trajectory optimization.}

\begin{figure}[!tbp]
	\centering
	\subfloat[Trajectory in the horizontal plane for different intializations. ]{\includegraphics[width=0.6\textwidth]{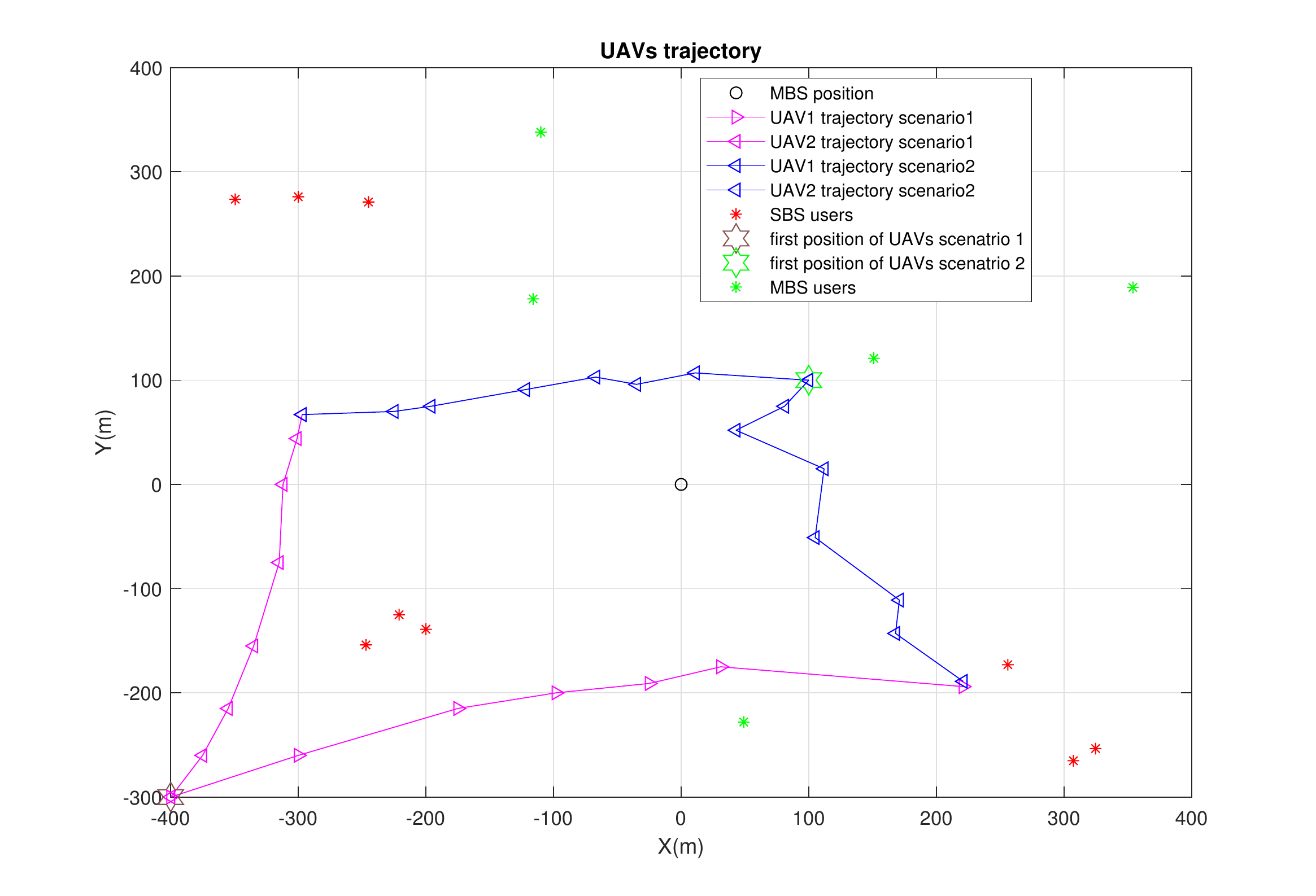}\label{fig:D2}}
	\hfill
	\subfloat[Trajectory in the 3D plane for different initializations.]{\includegraphics[width=0.6\textwidth]{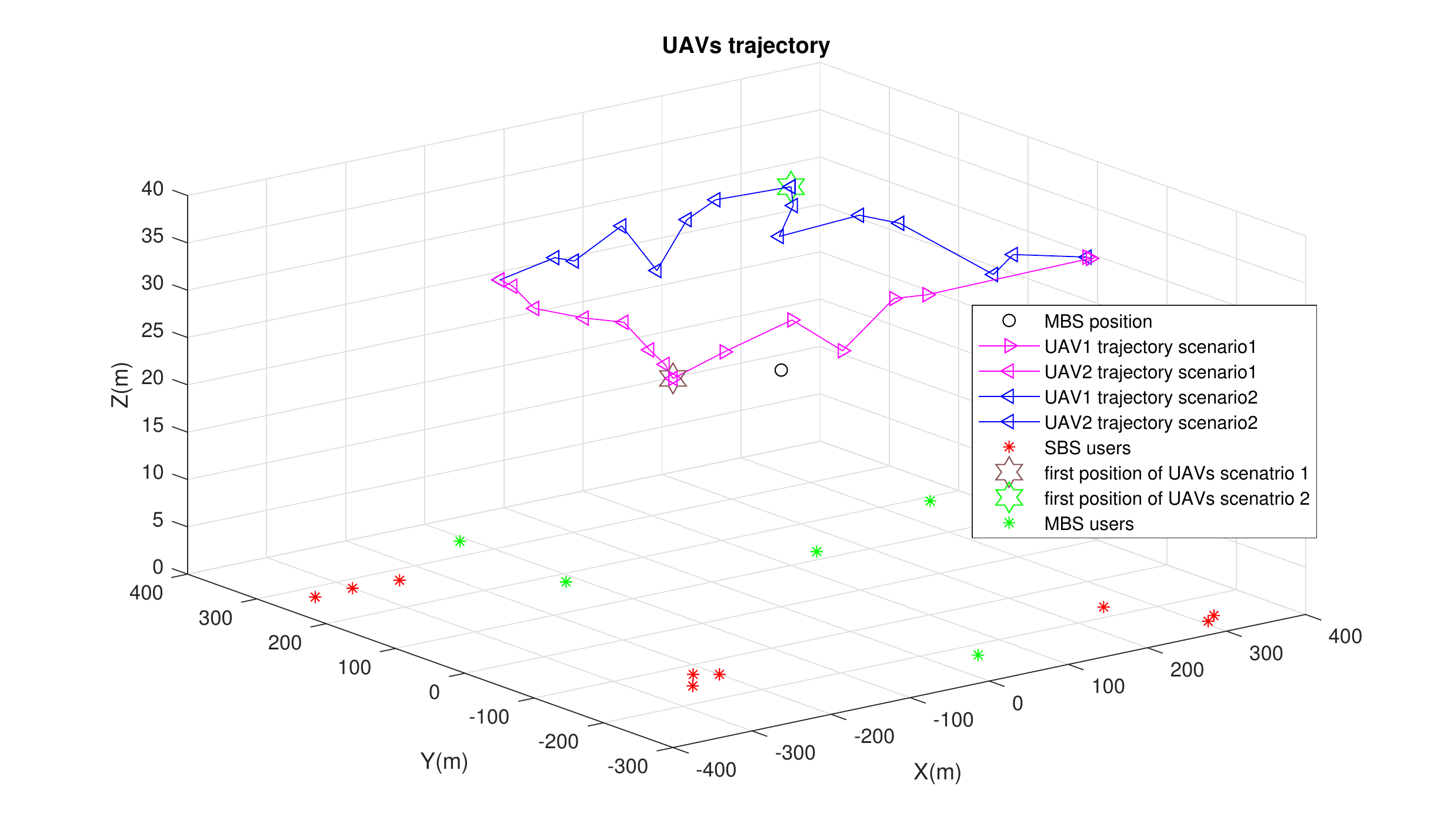}\label{fig:D3}}
	\caption{\textcolor{blue}{\small UAVs' trajectories of the proposed scheme.}}
\end{figure}

\subsection{Transmission power and Running time vs Number of RISs and Number of UAVs}
\begin{figure}[t]
	\centering
	\subfloat[Transmission power and running time versus the different number of reflecting elements at the RIS with one UAV. ]{\includegraphics[width=0.45\textwidth]{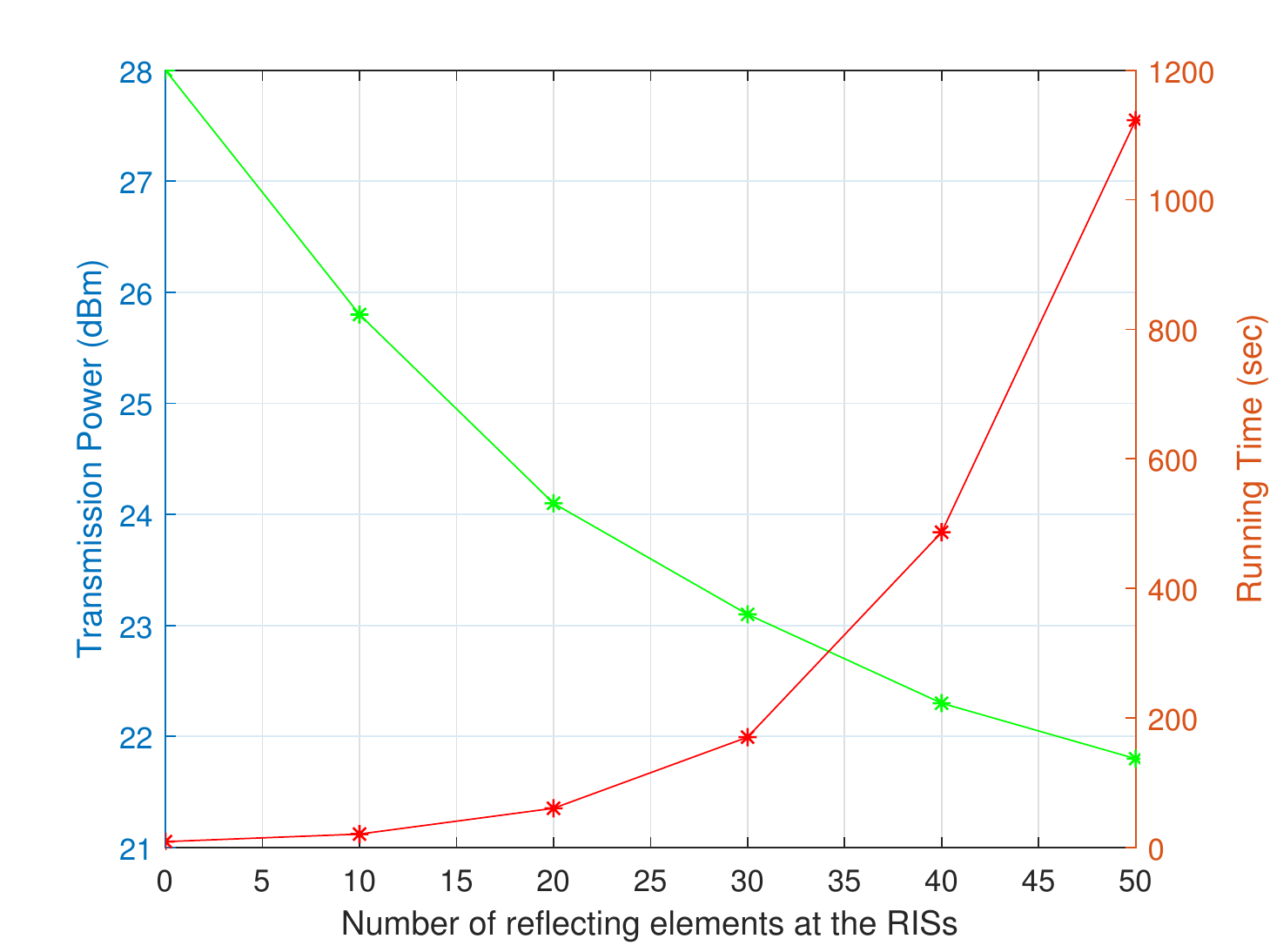}\label{fig:DD2}}
	\hfill
	\subfloat[Transmission power and running time versus the different number of UAVs with $N=10$.]{\includegraphics[width=0.45\textwidth]{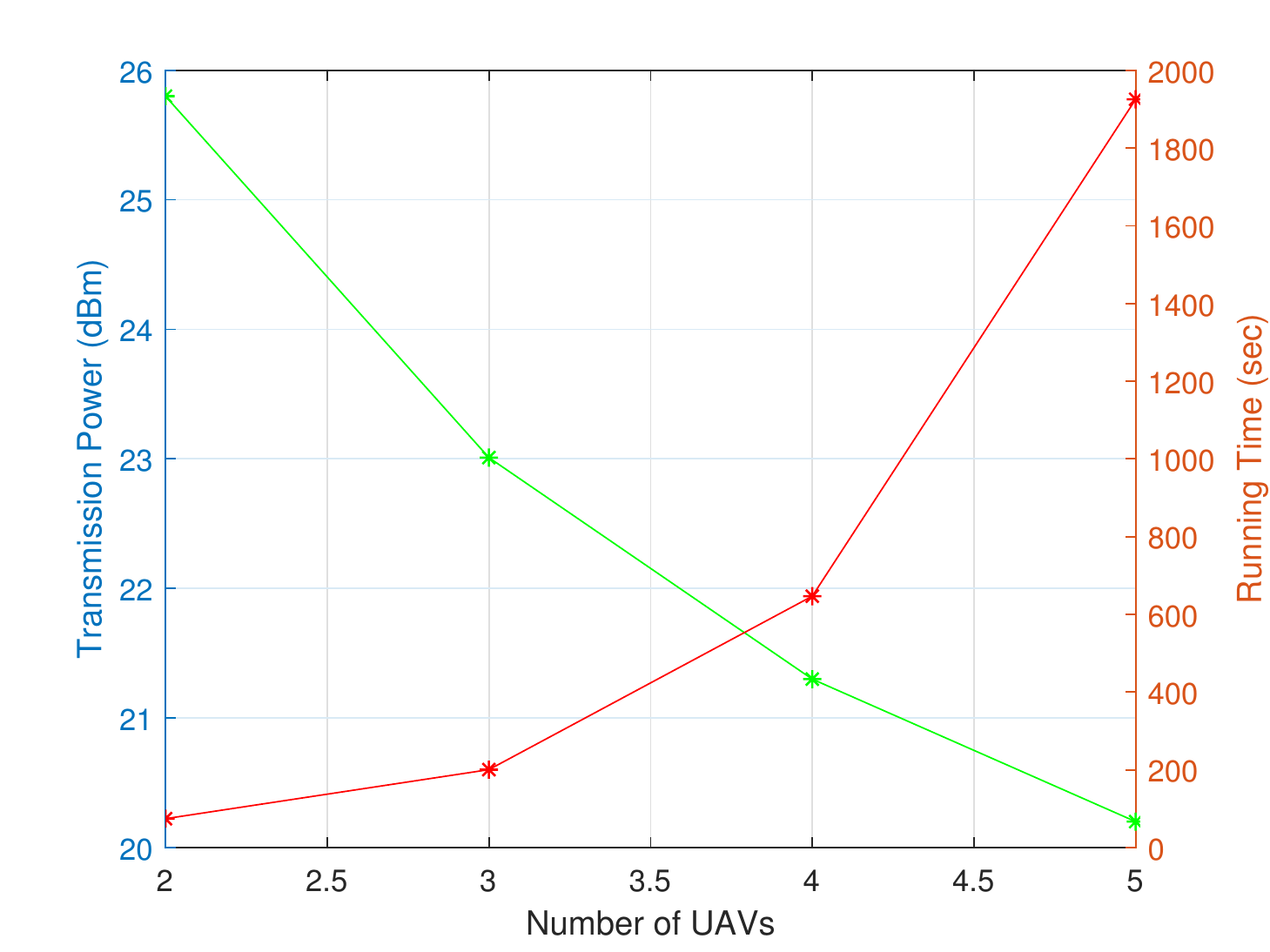}\label{fig:DD3}}
	\caption{\small \textcolor{blue}{Transmission power and running time.}}
\end{figure}

\textcolor{blue}{Fig. \ref{fig:DD2} and Fig. \ref{fig:DD3} show transmit power versus the number of reflecting elements at the RIS and number of UAVs, respectively. In both figures, by increasing the number of reflecting elements and number of UAVs, transmit power decrease significantly. For instance, by changing the number of UAVs from $3$ to $4$ with $N=10$, we observe $2$ dBm decrease in transmit power. On the other hand, by using only on UAV and varying the number of reflecting elements from $10$ to $30$, $4$ dBm can be saved in transmit power. This figure also shows the running time versus the reflecting elements at the RIS and number of UAVs, respectively. It can be seen that by increasing the number of reflecting elements and number of UAVs, the running time starts to increase and grow significantly for the higher number of reflecting elements and number of UAVs. These observations confirm that by increasing the number of reflecting elements, the resource allocator can decrease the transmit power with much lower complexity. However, by increasing the number of UAVs, the transmit power is smaller compared to the case where the number of reflecting elements increases since more DoF are available because of the trajectory of the UAVs. From Fig. 9a and Fig. 9b, one can conclude that by adding more UAVs, although the LoS links are established to facilitate communication, the proposed algorithm takes long time to converge. In fact, there is a tradeoff between complexity (running time) and performance gain.}
%\vspace{-5mm}
\section{Conclusions}\label{conclusion}
\textcolor{blue}{In this paper, we proposed dueling DQN learning along with classical optimization-based approaches to enable efficient deployment of multiple UAVs-RISs serving multiple DL users with the aid of the MC-PD-NOMA scheme. In particular, the total transmit power minimization problem was formulated by taking into account the maximum transmit power budget, minimum rate requirement of each user, and RIS as well as UAV limitations by jointly optimizing UAVs's trajectories/velocities, subcarrier allocations, phase shifts at each RIS, and active beamformers at the MBS and the SBSs. To solve this complicated problem, the main problem was divided into two sub-problems: the trajectories/velocities of the UAVs, phase shifts at the RISs, and subcarrier allocations for the mmW are solved by the learning approach via dueling DQN method, while joint beamforming design and subcarrier allocation for the $\mu$W subproblem are handled by a classical optimization problem. Simulation results exhibit the effectiveness of the proposed resource allocation scheme for the considered network with multiple UAVs-RISs under the MC-PD-NOMA scheme. They also reveal their superior performance in terms of the total transmit power compared to other baseline schemes. In particular, by deploying UAVs-RISs, the transmit power can be reduced by $6$ dBm while maintaining similar guaranteed QoS.}

%\vspace{-8mm}

\end{document}